\documentclass[conference]{IEEEtran}

\usepackage[utf8]{inputenc}
\usepackage[T1]{fontenc}
\usepackage{textcomp}
\usepackage{mathptmx}

\usepackage[usenames,dvipsnames]{xcolor}
\usepackage[hidelinks]{hyperref}

\usepackage{pgfplots}
\usepackage{pgfplotstable}
\usepackage{filecontents}
\usepgfplotslibrary{units}
\usetikzlibrary{calc, spy, patterns, backgrounds}

\pgfplotsset{
	my ybar legend/.style={
		legend image code/.code={
			\draw [##1] (0cm,-0.6ex) rectangle +(1.75em,1.1ex);
		},
	},
}

\usepackage{booktabs}

\usepackage{graphicx, caption, subcaption}
\usepackage{acro}

\usepackage{relsize}
\usepackage[shrink=20]{microtype}
\usepackage{csquotes}
\usepackage{listings}
\usepackage{algpseudocode}
\usepackage[curves]{struktex}
\usepackage{mathrsfs}
\usepackage{multicol}

\usepackage{acro}
\newcommand{\acrodef}[2]{\DeclareAcronym{#1}{short={#1},long={#2}}}
\acrodef{AABB}{axis-aligned bounding box}
\acrodef{API}{application programming interface}
\acrodef{ASIC}{application specific integrated circuit}
\acrodef{AST}{abstract syntax tree}
\acrodef{BRAM}{block RAM}
\acrodef{CER}{communication-to-execution ratio}
\acrodef{CG}{conjugate gradient}
\acrodef{CPI}{cycles per instruction}
\acrodef{CPU}{central processing unit}
\acrodef{CUDA}{compute unified device architecture}
\acrodef{CST}{concrete syntax tree}
\acrodef{DPM}{delay propagation mechanism}
\acrodef{DOM}{delay overlapping mechanism}
\acrodef{DPOM}{delay propagation and overlapping mechanisms}
\acrodef{DSL}{domain-specific language}
\acrodef{DVFS}{dynmic voltage frequency scaling}
\acrodef{FD}{finite difference}
\acrodef{FEM}{finite element method}
\acrodef{FFC}{FEniCS Form Compiler}
\acrodef{FFT}{Fast Fourier transform}
\acrodef{FIFO}{first in first out}
\acrodef{FLOPS}{floating point operations per second}
\acrodef{FPGA}{field-programmable gate array}
\acrodef{FV}{finite volume}
\acrodef{GMRES}{generalized minimal residual}
\acrodef{GPU}{graphics processor unit}
\acrodef{GS}{Gauss-Seidel}
\acrodef{GUI}{graphical user interface}
\acrodef{HDL}{hardware description language}
\acrodef{HHG}{hierarchical hybrid grid}
\acrodef{HLS}{high-level synthesis}
\acrodef{HPC}{high-performance computing}
\acrodef{IP}{intellectual property}
\acrodef{ITAC}{Intel trace analyzer and collector}
\acrodef{IR}{intermediate representation}
\acrodef{JIT}{just-in-time}
\acrodef{LFA}{local Fourier analysis}
\acrodef{LBM}{Lattice Boltzmann}
\acrodef{LoC}{lines of code}
\acrodef{LZR}{Leibniz Supercomputing Centre}
\acrodef{MPI}{Message Passing Interface}
\acrodef{NDG}{nodal discontinuous Galerkin}
\acrodef{NDGTD}{nodal discontinuous Galerkin time domain}
\acrodef{NIC}{network interface controller}
\acrodef{OMP}{OpenMP}
\acrodef{OS}{operating system}
\acrodef{P2P}{point-to-point}
\acrodef{PDE}{partial differential equation}
\acrodef{PGAS}{partitioned global address space}
\acrodef{PPnR}{post place and route}
\acrodef{QDR}{quad data rate}
\acrodef{RAM}{random access memory}\acuse{RAM}
\acrodef{RBGS}{red-black Gauss-Seidel}
\acrodef{RDMA}{remote direct memory access}
\acrodef{RHS}{right-hand side}
\acrodef{RRZE}{Regional Computer Center Erlangen} 
\acrodef{RTL}{register transfer level}
\acrodef{SHM}{shared memory}
\acrodef{SPIR}{standard portable intermediate representation}
\acrodef{SPL}{software product lines}
\acrodef{SIMD}{single instruction, multiple data}
\acrodef{SMP}{symmetric multiprocessing}
\acrodef{SMT}{simultaneous multithreading}
\acrodef{STL}{Standard Template Library}
\acrodef{TLB}{translation lookaside buffer}
\acrodef{TPDL}{target platform description language}
\acrodef{XML}{eXtensible Markup Language}

\usepackage{amsmath}
\usepackage{amsmath}
\usepackage{amsfonts}

\usepackage{relsize}
\usepackage{xspace}
\newcommand{\CPP}{C\nolinebreak[4]\hspace{-.05em}\raisebox{.23ex}{\relsize{-1}{++}}}

\newif\iftitle
\titletrue

\newcommand{\bq}{\begin{equation}}
\newcommand{\eq}{\end{equation}}

\newcommand{\byte}{\mbox{byte}}
\newcommand{\second}{\mbox{s}}

\newcommand{\MuS}{\mbox{$\mu$s}}

\newcommand{\flop}{\mbox{flop}}

\newcommand{\bit}{\mbox{bit}}

\newcommand{\FS}{\mbox{\flop/\second}}

\newcommand{\GBS}{\mbox{GB/\second}}

\newcommand{\GHZ}{\mbox{GHz}}
\newcommand{\ns}{\mbox{ns}}

\newcommand{\GB}{\mbox{GB}}

\newcommand{\MB}{\mbox{MB}}

\newcommand{\eos}{~.}
\newcommand{\cma}{~,}

\acuse{CPU}
\acuse{GPU}
\usepackage{todonotes}
\usepackage{setspace}
\makeatletter
\renewcommand{\todo}[2][]{\@todo[caption={#2},#1]{\begin{spacing}{0.5}\fontfamily{phv}\fontseries{mc}\selectfont{#2\vspace{-1em}}\end{spacing}}}
\makeatother 

\colorlet{gwcolor}{Green} % Christian
\colorlet{ghcolor}{ProcessBlue}
\colorlet{aycolor}{RubineRed}
\usepackage[binary-units=true]{siunitx}

\usepackage{cleveref}

\newif\ifblind
\blindfalse

\makeatletter
\newcommand{\verbatimfont}[1]{\renewcommand{\verbatim@font}{\ttfamily#1}}
\makeatother

\begin{document}
\title{Propagation and Decay of Injected One-Off Delays on Clusters: A Case Study}
%\title{\huge Delay Propagation and Overlapping Mechanisms on Clusters\\ A Case Study of Idle Period based on Workload, Communication, and Delay Granularity}
%\title{DOM: Delay Oscillation Modelling \\ A Case Study of Delays based on Configurable Parameters Space}

\ifblind
\author{Authors omitted for double-blind review process}
\else
\author{
\IEEEauthorblockN{Ayesha Afzal, Georg Hager, Gerhard Wellein}
\IEEEauthorblockA{Erlangen Regional Computing Center, Friedrich-Alexander University Erlangen-N\"urnberg, Germany} %, {ayesha.afzal, georg.hager, gerhard.wellein}@fau.de}
}
\fi

\maketitle
\pgfkeys{/pgf/number format/.cd,1000 sep={\,}}
\begin{abstract}
Analytic, first-principles performance modeling of distributed-memory
applications is difficult due to a wide spectrum of random
disturbances caused by the application and the system.  These
disturbances (commonly called \enquote{noise}) destroy the assumptions
of regularity that one usually employs when constructing simple
analytic models.  Despite numerous efforts to quantify, categorize,
and reduce such effects, a comprehensive quantitative understanding of
their performance impact is not available, especially for
long delays that have global consequences for the parallel application.
In this work, we
investigate various traces collected from synthetic
benchmarks that mimic real applications on simulated and real message-passing
systems in order to pinpoint the mechanisms behind delay propagation.
We analyze the dependence of the propagation speed of idle
waves emanating from injected delays
with respect to the execution and communication properties of the
application, study how such delays decay under increased noise levels,
and how they interact with each other.
We also show how fine-grained noise can make a system immune
against the adverse effects of propagating idle waves.
Our results contribute to a better understanding of the
collective phenomena that manifest themselves in distributed-memory
parallel applications.
\end{abstract}

\section{Introduction}

\subsection{Simple analytic models and noise}

White-box analytic performance models of parallel applications are built on
simplifying assumptions about the interactions of code with the hardware. For
example, the Roof{}line model~\cite{roofline:2009} predicts the performance of
parallel loops on multicore processors by assuming that data transfers to and
from main memory overlap perfectly with the execution of work in the cores, and
whichever takes longer determines the runtime. Distributed-memory parallel
applications require some communication model in addition, such as the Hockney
model~\cite{Hockney:1994}, LogP~\cite{Culler:1993}, or one of their variants
and extensions.
Further assuming a bulk-synchronous model of computation without overlap
of communication and computation, the overall runtime
of a program may then be predicted by adding the execution time to the
communication time: $T=T_\mathrm{exec}+T_{comm}$\@. Many refinements
are possible, and even though the construction
of analytic models is often tedious, they lead to invaluable insights
about bottlenecks and governing mechanisms.

Many of the assumptions underlying those models are fulfilled only approximately
in practice, which limits their accuracy and, if deviations become too large,
their usefulness. In this paper, we want to shed light on a phenomenon that
violates the bulk-synchronous assumption about a parallel program on typical
cluster systems comprising many nodes with multicore processors: Even if the
workload is perfectly balanced across all workers, slight variations in
execution performance or communication time can lead to prominent collective
phenomena that break the regular ``lockstep'' pattern. Such variations exist in
all real computer systems, and they have a wide spectrum of origins.  A very
coarse categorization of statistical causes for variation can be made:
\emph{Fine-grained noise} originates from OS interference, clock speed
fluctuations, interrupts, data management in driver software, and more, and
usually leads to delays of the order of microseconds. \emph{Execution delays}
are longer and more coarse-grained; examples are regular administrative jobs
(e.g., cron job scripts), page faults, garbage collection, or infrequent
application imbalances that take a significant amount of a core's resources
for a long time, maybe longer than the typical periodicity of a bulk-synchronous
program. We will keep the distinction between \emph{noise} and \emph{delays} in
this work.

\subsection{Motivating examples}
\begin{figure*}[tb]
	\centering
	 \begin{subfigure}[t]{.6\textwidth}
	\begin{tikzpicture}
		\pgfplotstableread{figures/Pi_TRIAD_strongscaling/BlockingTRIAD_50MB_2MB_1000t.txt}\mydataZ;
		\pgfplotstableread{figures/Pi_TRIAD_strongscaling/BlockingTRIAD_50MB_2MB_1000t_Socket.txt}\mydataS;
		\pgfplotstableread{figures/Pi_TRIAD_strongscaling/BlockingTRIAD_50MB_2MB_1000t_total.txt}\mydataZZ;
		\pgfplotstableread{figures/Pi_TRIAD_strongscaling/BlockingTRIAD_50MB_2MB_1000t_Socket_total.txt}\mydataSS;
		\pgfplotstableread{figures/Pi_TRIAD_strongscaling/model.txt}\myModel;
		\pgfplotstableread{figures/Pi_TRIAD_strongscaling/model_total.txt}\myModell;
		\begin{axis}[trim axis left, trim axis right, scale only axis,
			width = .48\columnwidth,
			height = 0.22\textheight,
			ymin=0,
			xlabel = {Number of sockets},
			ylabel = {Performance [\si{\giga F \per \second}]},
			y label style={at={(0.06,0.5)}},
			legend style = {font=\footnotesize, nodes={inner sep=0.03em}, anchor=south, at={(0.5,1.05)}},
			legend columns = 1,
			 name=ax1
			]
	
			\addplot[mark=diamond*, only marks, style={blue, fill=none}, error bars/.cd, y dir=both, y explicit,]
			table
			[
			x expr=\thisrow{NS}, 
			y error minus expr=\thisrow{Avg}-\thisrow{Min},
			y error plus expr=\thisrow{Max}-\thisrow{Avg},
			]{\mydataS};  
			
			\addplot[mark=square*, only marks,  style={blue, fill=none}, error bars/.cd, y dir=both, y explicit,]
			table
			[
			x expr=\thisrow{NS}, 
			y error minus expr=\thisrow{Avg}-\thisrow{Min},
			y error plus expr=\thisrow{Max}-\thisrow{Avg},
			]{\mydataSS};
				
%			\addplot [ybar,bar width=2pt,style={blue},pattern=horizontal lines, my ybar legend]
%			table
%			[ 
%			x expr=\thisrowno{0},  
%			y expr=\thisrowno{3}
%			]{\mydata};
%			
			\addplot [mark=diamond*, only marks, style={Bittersweet, fill=Bittersweet}]
			table
			[
			x expr=\thisrow{NS},
			y expr=\thisrow{Model},
			]{\myModel};
			
			\addplot [mark=square*, only marks, style={Bittersweet, fill=Bittersweet}]
			table
			[
			x expr=\thisrow{NS},
			y expr=\thisrow{Model},
			]{\myModell};

			\legend{Execution performance (PPN=20), Total performance (PPN=20), Execution model (PPN=20), Total model (PPN=20)} 
			\coordinate (pt) at (axis cs:0.4,8);
			\coordinate (c2) at (axis cs:1.1,32);
			\draw (pt) rectangle (axis cs:2.2,0.01);
			\end{axis}
%			\node[pin=87:{%
%			\begin{tikzpicture}[baseline,trim axis left,trim axis right]
			\begin{axis}[		
			   name=ax2,
			 at={($(ax1.south east)+(0.5cm,0.35cm)$)},
%			anchor=north west,
			width = .5\columnwidth,
			height = 0.265\textheight,
%			small,
			xlabel = {Number of processes},
			x label style={at={(0.5,0)}}, %overlay, 
			xmin=0,xmax=21,
			ymin=0,ymax=7.5,
%			enlargelimits,
			]
			\addplot[mark=diamond*, only marks,  style={blue, fill=none}, restrict x to domain=1:20, error bars/.cd, y dir=both, y explicit,]
			table
			[
			x expr=\thisrow{NP}, 
			y error minus expr=\thisrow{Avg}-\thisrow{Min},
			y error plus expr=\thisrow{Max}-\thisrow{Avg},
			]{\mydataZ};
			
			\addplot[mark=square*, only marks,  style={blue, fill=none}, restrict x to domain=1:20, error bars/.cd, y dir=both, y explicit,]
			table
			[
			x expr=\thisrow{NP}, 
			y error minus expr=\thisrow{Avg}-\thisrow{Min},
			y error plus expr=\thisrow{Max}-\thisrow{Avg},
			]{\mydataZZ};
			
			\addplot [mark=diamond*, only marks, style={Bittersweet, fill=Bittersweet}, restrict x to domain=1:20] coordinates {(10,2.5) (20,5) };
			
			\addplot [mark=square*, only marks, style={Bittersweet, fill=Bittersweet}, restrict x to domain=1:20] coordinates {(10	,2.419354839
) (20,	4.6875) };
			
			\end{axis}
			\draw [dashed] (pt) -- (ax2.north west);
			\draw [dashed] (1.1,0) -- (ax2.south east);
			\node [font=\small] at (2.5,-1.3){(a)}; 
			\node [font=\small] at (7.5,-1){(b)}; 
			\end{tikzpicture}%
%		}] at (c2) {};
%		\end{tikzpicture} 
	\end{subfigure}
	\begin{subfigure}[t]{.39\textwidth}
		\begin{tikzpicture}
		\pgfplotstableread{figures/Pi_TRIAD_strongscaling/BlockingTRIAD_50MB_2MB_1000t_1PPN.txt}\mydataN;
		\begin{axis}[trim axis left, trim axis right, scale only axis,
		width = .75\columnwidth,
		height = 0.22\textheight,
		ymin=0,
		xlabel = {Number of nodes},
		ylabel = {Performance [\si{\giga F \per \second}]},
		y label style={at={(0.1,0.5)}},
		legend style = {font=\footnotesize, nodes={inner sep=0.03em}, anchor=south, at={(0.5,1.05)}},
		legend columns = 1,
		]
		
		\addplot+[mark=square*, only marks, error bars/.cd, y dir=both, y explicit,]
		table
		[
		x expr=\thisrow{NN}, 
		y error minus expr=\thisrow{AVG}-\thisrow{MIN},
		y error plus expr=\thisrow{MAX}-\thisrow{AVG},
		]{\mydataN};
		
		\addplot+ [mark=square*, only marks, style={Bittersweet, fill=Bittersweet}]
		table
		[
		x expr=\thisrow{NN},
		y expr=\thisrow{MODEL},
		]{\mydataN};
		\legend{Total Performance (PPN=1), Total model (PPN=1)} 
		\end{axis}
		\node [font=\small] at (2.5,-1.3){(c)}; 
		\end{tikzpicture}
	\end{subfigure}	
	\caption{Comparison of actual measurements with a performance model for
          strong scaling of an MPI-parallel STREAM triad benchmark on a cluster
          system with 20 cores per node and InfiniBand interconnect. See
          main text for details of the setup and model. (a) Total
          measured (blue squares) and model (red squares) performance on up to 9
          full sockets, execution-only performance model
          (red diamonds), and execution-only median and min/max
          performance across all cores (blue diamonds and whiskers).
          (b) Closeup of the node level. (c) Comparison
          of model and measurement when running only one core per node.
        }
	\label{fig:TRIAD}
\end{figure*}
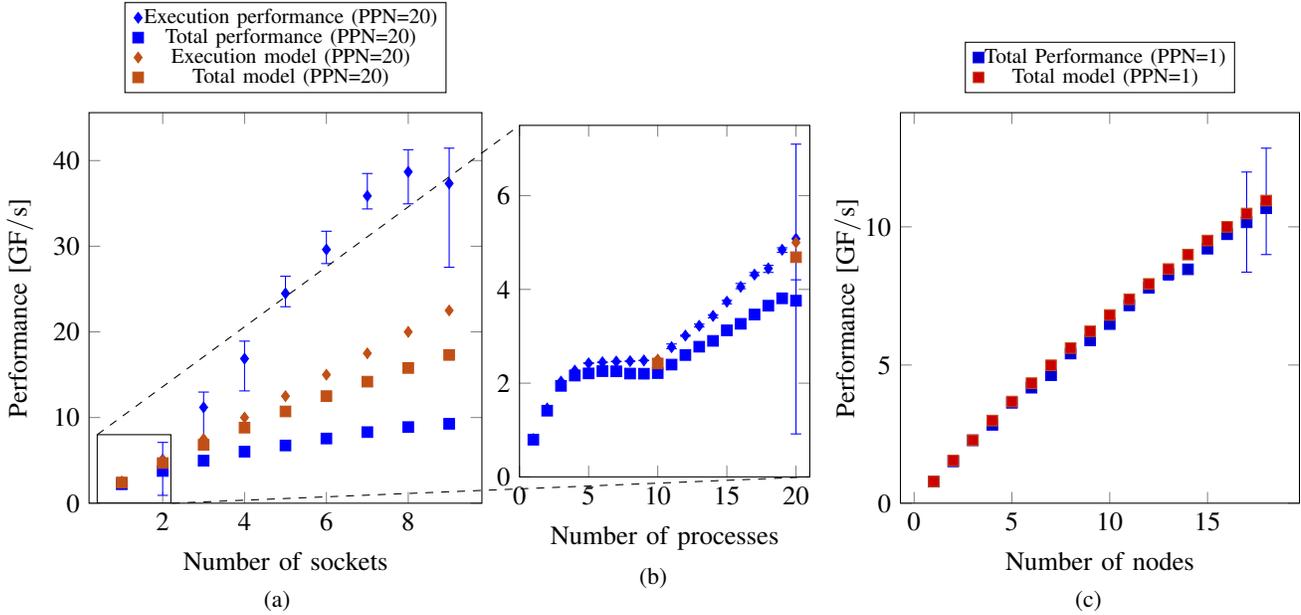
One
prominent result of noise- or delay-induced desynchronization is that
simple execution/communication performance models as described above
tend to deliver very inaccurate predictions, and those can be
optimistic as well as pessimistic. We performed a simple experiment that
shows this effect (see Figure~\ref{fig:TRIAD}): On an InfiniBand cluster
with dual-socket Intel ``Ivy Bridge'' nodes (details on the hardware
can be found in Section~\ref{sec:Emmy}), we ran a pure-MPI version of the
McCalpin STREAM triad~\cite{mccalpin1995memory} loop (\verb.A(:)=B(:)+s*C(:).)
in a strong scaling
scenario. An overall working set of $V_\mathrm{mem}=1.2\,\GB$ ($5\times10^7$ elements)
is split evenly across the MPI processes. After each full loop
traversal, each MPI rank $i$ sends and receives $V_\mathrm{net}=2\,\MB$ of data to
and from each
of its direct neighbors $i+1$ and $i-1$ (the process topology is a closed ring).
All nodes are connected to a single, fully nonblocking
IB leaf switch to eliminate interference from other cluster jobs
and enable full non-blocking bandwidth. Given the memory bandwidth
of a socket ($b_\mathrm{mem}\approx40\,\GBS$) and the asymptotic node-to-node
bandwidth of the InfiniBand network ($b_\mathrm{net}\approx 3\,\GBS$) one can construct a
simple optimistic model for the runtime of one compute-communicate
cycle. When $n$ is the number of sockets (each running ten processes),
the predicted time per time step is
\bq\label{eq:simplemodel}
T(n)=\frac{V_\mathrm{mem}}{nb_\mathrm{mem}}+\frac{2V_\mathrm{net}}{b_\mathrm{net}}\eos
\eq
Here we ignore the communication between processes within a node,
(which can be significant in practice, but neglecting it makes the model
just more optimistic). The performance in \FS\ is then just
$P(n)=2\cdot5\times10^7\,\flop/T(n)$. 
Figure~\ref{fig:TRIAD}(a) shows
the total predicted (red squares) and measured (blue squares)
performance
versus the number of sockets. The strong deviation of almost
a factor of two at 9 nodes might be expected because of our
ignorance of intra-node communication; it is surprising,
however, that the pure execution performance (calculated
from the mean execution time of individual processes)
is so much higher than the prediction, which assumes
linear scalability, of course (blue vs.\ red diamonds).
Obviously, the
assumption that computation and communication do not overlap cannot
be true, but since the load is perfectly balanced across processes,
this overlap must be something that emerges automatically during
the program's runtime due to the intrinsic system noise. Indeed,
a trace analysis reveals that the MPI processes are massively out
of sync, which leads to automatic overlap and mitigation of the
memory bandwidth bottleneck but also causes massive delays
due to processes waiting for messages. Due to the statistical
nature of this effect, 
variations across runs and processes are inevitable, as
shown by the min/max whiskers. Figure~\ref{fig:TRIAD}(b)
is a zoom-in on the node level, where the simple bandwidth
model works fine on up to one socket but communication overhead
becomes visible beyond. 
In Figure~\ref{fig:TRIAD}(c)
we show a similar  experiment with only one process per node.
The relative communication overhead is now much smaller since
the node-level performance is only about 1/6 of the saturated case.
With less opportunity for overlap, and the memory bandwidth
bottleneck removed, the model actually delivers a good
prediction of the average performance although some
outliers at the larger node counts indicate the onset of
overlap.

\begin{figure*}[tb]
	\centering
	\includegraphics[width=\textwidth]{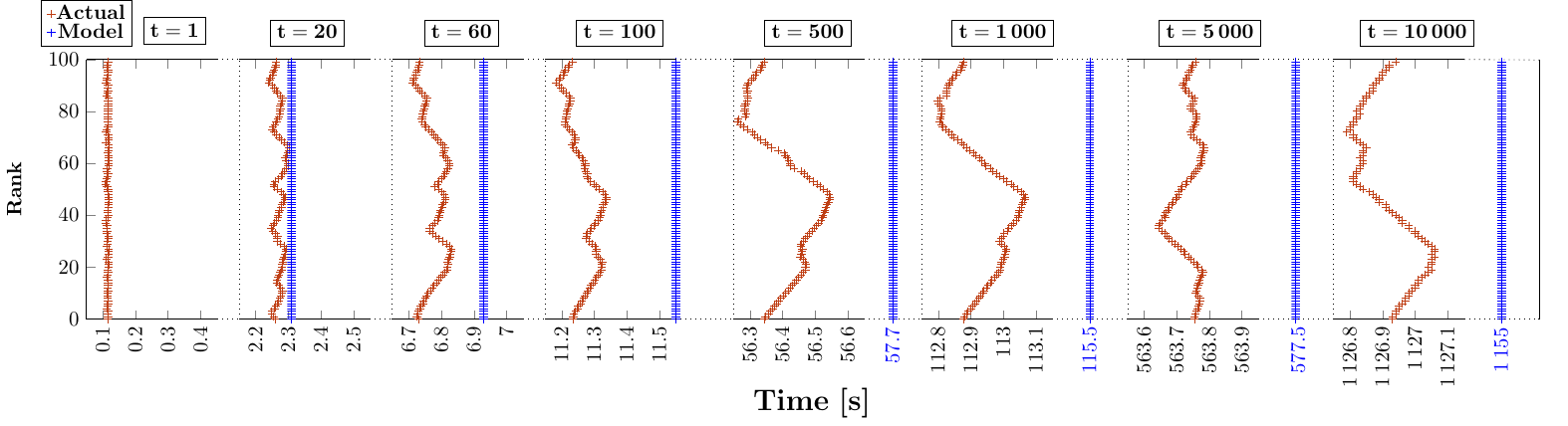}
\caption{Measured LBM irregular structure in comparison with expected model regularity. The plot shows that the time steps with more irregularity (iterations in between and at the end of the results) manifest maximum percentage variation towards better performance ($2.25$\%  $@$t$500$, t$1\,000$ and $2.42$\% $@$t$10\,000$).} 
\label{fig:LBM}
\end{figure*}
Beyond simple toy benchmarks, real-world applications show similar patterns of
nonsynchronicity. On the same system as above we ran an MPI-parallel double
precision Lattice-Boltzmann (LBM) fluid solver with D3Q19 discretization
and a single relaxation time (SRT) model on five nodes
(100 cores). The node-level performance properties of such codes were
thoroughly investigated in~\cite{Wittmann:2016}. We use an overall problem size
of $302^3$ lattice cells (including a boundary layer in each diraction)
for a working set of more than $8\,\GB$. Domain
decomposition is done only along the outer dimension with periodic boundary
conditions, leading
to a rather large communication overhead of at least 30\% of the runtime.
In Figure~\ref{fig:LBM} we show
timeline snapshots at different LBM time steps $t$, from $t=1$ to $t=10,000$, where
the location of the time step along the wall-clock time axis is marked on each
process (red markers). For reference, we also show the expected positions
according to the simple nonoverlapping execution-communication model
(\ref{eq:simplemodel}). While the deviation from the model and the variation
across processes is still small after 20 time steps ($<0.1\,\mbox{s}$), a global structure
has emerged at $t=500$ with a fundamental ``wavelength'' equal to the size
of the system (100 processes) and an amplitude of 0.3 wall-clock seconds.
This pattern is not static but shifts and changes shape, as can be seen
to $t=5,000$ and $t=10,000$. Moreover, the actual runtime at $t=10,000$ is
about 28\,s smaller than expected. While this is only a deviation of
about 2.5\%, the pattern is interesting and may show up more prominently
with applications that have different communication overheads and
patterns.

The examples above have demonstrated that some scenarios allow noise to act as
an application accelerator as well as a slowdown factor. There is, however, a
very complex interplay between application code execution, the message passing
library, and the network, which leads to a rich spectrum of local and collective
phenomena in parallel code, especially when noise is present. The accelerating
effect is certainly not guaranteed. In this work we want to study a particular
aspect of this theme: the wave-like propagation of execution delays (``idle
waves'') through the network under the influence of system noise and variable
injected noise.

\subsection{Contributions}

The major contribution of this paper is the investigation of
idle waves, which emanate from strong delays occurring on
individual processes of an MPI-parallel program.
\begin{itemize}
\item We investigate and categorize the mechanisms of the propagation of ``idle
  waves'' emanating from execution delays across communicating processes under
  some simplifying assumptions, notably a bulk-synchronous application
  structure.
\item We show how such idle waves interact and (partially)
  cancel each other, proving that a linear wave equation is
  inappropriate to describe the phenomenon.
\item We give an analytic expression for the speed of an idle wave
  in a noise-free homogeneous system under core-bound computational
  load, taking into account execution time,
  communication time, communication topology, and communication mode
  (eager vs.\ rendezvous).
\item We investigate the impact of injected, fine-grain exponential
  noise on the propagation speed and lifetime of idle waves
  and show how the decay of the wave depends on the strength of the
  noise.
\item We demonstrate that the application slowdowm usually
  caused by strong idle waves may be unobservable due to
  the presence of noise. 
\end{itemize}
This paper is organized as follows: In Section~\ref{sec:Categ} we introduce some
important terms and categorize the execution and communication
scenarios under investigation. Section~\ref{sec:Testbed} gives details about
our hardware and software setup and the inherent node-level noise structure
of the cluster system we use for the benchmarks. The mechanisms of
delay propagation under various conditions are covered in Section~\ref{sec:Mechanics},
while Section~\ref{sec:Decay} deals with the analysis of idle waves decaying
under noise. Related work is covered in
Section~\ref{sec:Relatedwork}, and Section\ref{sec:conclusions} concludes the paper and
gives an outlook to future work.

\section{Categorization of parameters}\label{sec:Categ}

A multitude of system and application parameters and properties influence the
phenomenology of delay propagation and desynchronization.  This section tries
to categorize the most relevant factors.

\subsection{Node level: Scalability vs. saturation}

In \ac{HPC}, most parallel codes comprise sequences of back-to-back loops or
bags of tasks that manipulate data. This data is either already present in the
local memory hierarchy or must be communicated from elsewhere via message
passing.  Leaving aside communication for the moment, hardware-software
interaction on the compute node level can be categorized into \emph{data-bound}
and \emph{compute-bound} phases. Lacking any load imbalance, the performance of
truly compute-bound code will scale across the cores of the node since no shared
resources are on the critical path. In case of data boundedness, the actual data
transfer bottleneck may be a per-core (i.e., private) cache level, which will
not impede scalability; if a shared resource such as a shared cache or the
memory interface is involved, performance does not scale but saturate as the
number of cores increases. The motivating STREAM triad and LBM examples in the
introduction are clearly in the data-bound category.  With such a code, using
fewer than the maximum number of cores on the contention domain will usually not
change the performance, and some load imbalance in the form of a few
``speeders'' may be tolerated.  The Roof{}line~\cite{roofline:2009} and
Execution-Cache-Memory (ECM) \cite{sthw15,Hofmann:2018} performance models allow
a rather complete analytic prediction of steady-state loop performance on
multicore CPUs.

Note that manifest load
imbalance within a single phase is considered an application-induced delay here,
just as lock contention, false sharing, and similar effects (see below).

\subsection{System topology}

Clusters of dual-socket multicore nodes (with or without accelerators) are the
dominating high-performance parallel computer architecture today.  These systems
show a complex topology, in the sense that basic, identical components are
assembled on multiple levels to build a hierarchical structure: cores, chips,
ccNUMA domains, sockets, nodes, network islands. Apart from data bottlenecks on
the chip level (see above), which manifest themselves only when an on-chip
shared resource is exhausted, communication characteristics like latency
and asymptotic bandwidth for point-to-point and
collective primitives can be very different across intra-chip, inter-chip,
and inter-node connections.

\subsection{Communication features}\label{sec:Communication}

\subsubsection{Message-passing modes}\label{sec:Protocols}

Beyond the well-defined distinction between \emph{blocking} and \emph{nonblocking}
communication, most of the details about how communication takes place in
\ac{MPI} are left to the implementation. As a general rule, short messages sent
via the standard \verb.MPI_Send. are transferred using an \emph{eager protocol},
i.e., due to internal buffering there is no handshake between communicating
processes, which may entail ``automatic'' asynchronous message transfer. Larger
messages usually require a handshake (\emph{rendezvous protocol}), causing
synchronization and, probably, explicit, nonoverlapping message transfer.

The \ac{MPI} Standard is purposefully vague about how the eager protocol should
be implemented.  \ac{MPI} implementations often allow the user to choose the
protocol by setting an \enquote{eager limit} for different (intra-node and
inter-node) devices. This is an upper bound on the size of messages sent or
received using the eager protocol. Implementations also provide tuning knobs to
control the number and the size of shared memory buffers or other internal
parameters.  As a consequence, the performance gain of eager protocol over
rendezvous because of reduced synchronizing delays is also implementation
dependent.

\subsubsection{Communication patterns}\label{sec:Patterns}

Communication patters are highly application dependent, and we do not
attempt a full categorization here. Instead, we point out those patterns
that we will need for our experiments in later sections. We also
restrict ourselves to point-to-point communication patterns in this
work.

\paragraph{Unidirectional vs. bidirectional next-neighbor}

Although rarely seen in practice, unidirectional next-neighbor communication
along an ordered set of processes is a good starting point for studying
propagation phenomena.  Each process $P_\mathrm{i}$ receives data from one
neighbor process ($P_\mathrm{i+1}$ or $P_\mathrm{i-1}$) and sends it to the
other neighbor process ($P_\mathrm{i-1}$ or $P_\mathrm{i+1}$). In bidirectional
communication, each process $P_\mathrm{i}$ exchanges, i.e., both sends and
receives, data from its two neighbors $P_\mathrm{i \pm 1}$.

\paragraph{Next-neighbor vs. multiple-neighbor}

Generalizing on the previous pattern, each process can have multiple
neighbors on each direction. This occurs in many linear algebra
and domain decomposition scenarios and entails more rigid dependencies
across the processor grid.

\paragraph{Periodic vs. open boundaries}

If the process grid is nonperiodic, propagating disturbances die out at the
boundaries. Periodic boundary conditions (in one or more dimensions) enable the
propagation of disturbances over larger distances and allow for more
interactions between processes.

\section{Hardware testbed characterization}\label{sec:Testbed}

\subsection{Cluster systems and software}\label{sec:Emmy}

\begin{figure}[tb]
	\centering
  	\begin{subfigure}[t]{.5\textwidth}
  	\centering
	\includegraphics[width=\textwidth]{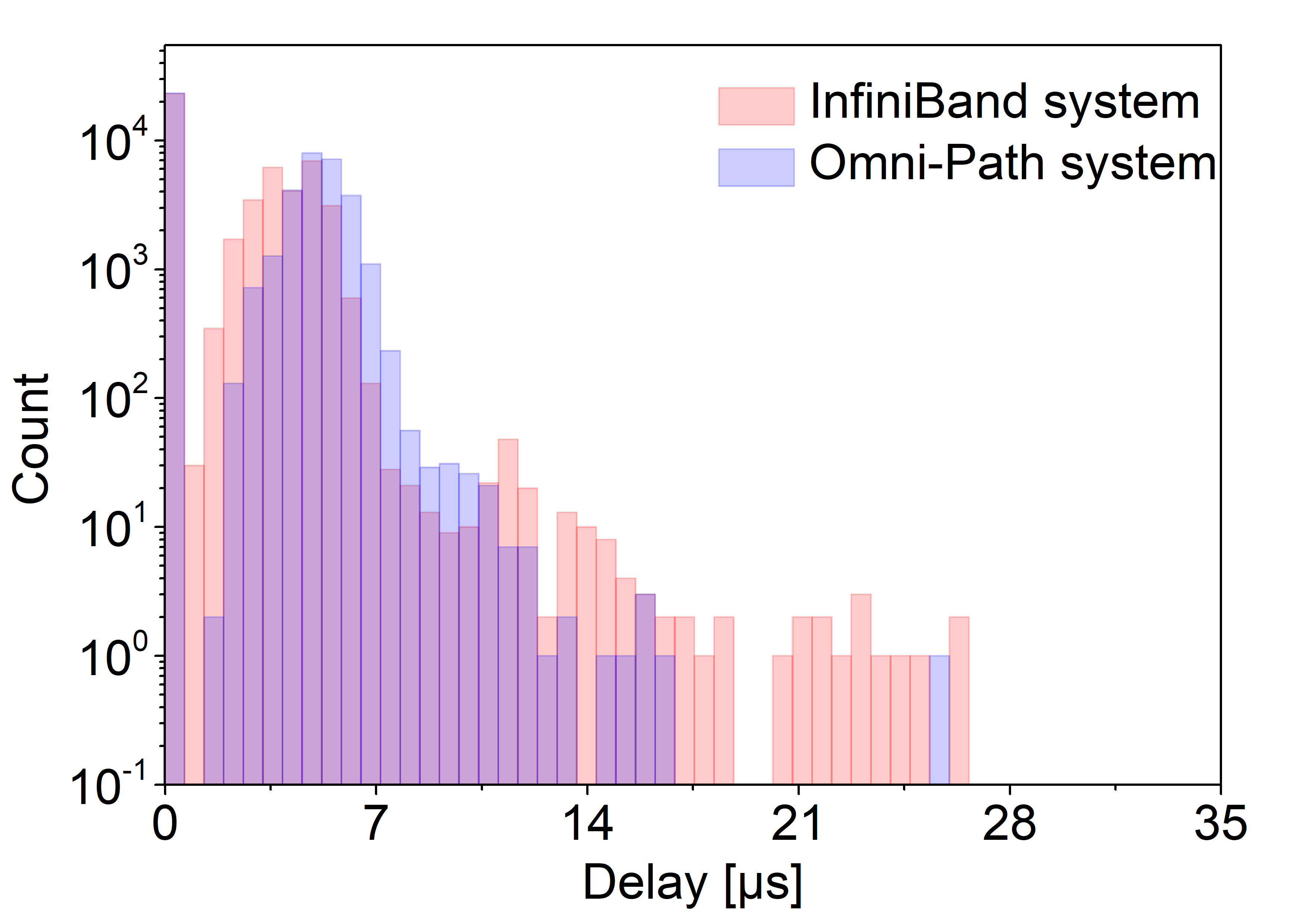}
	\caption{System noise with enabled SMT}
	\end{subfigure}
	\begin{subfigure}[t]{.5\textwidth}
	\centering
	\includegraphics[width=\textwidth]{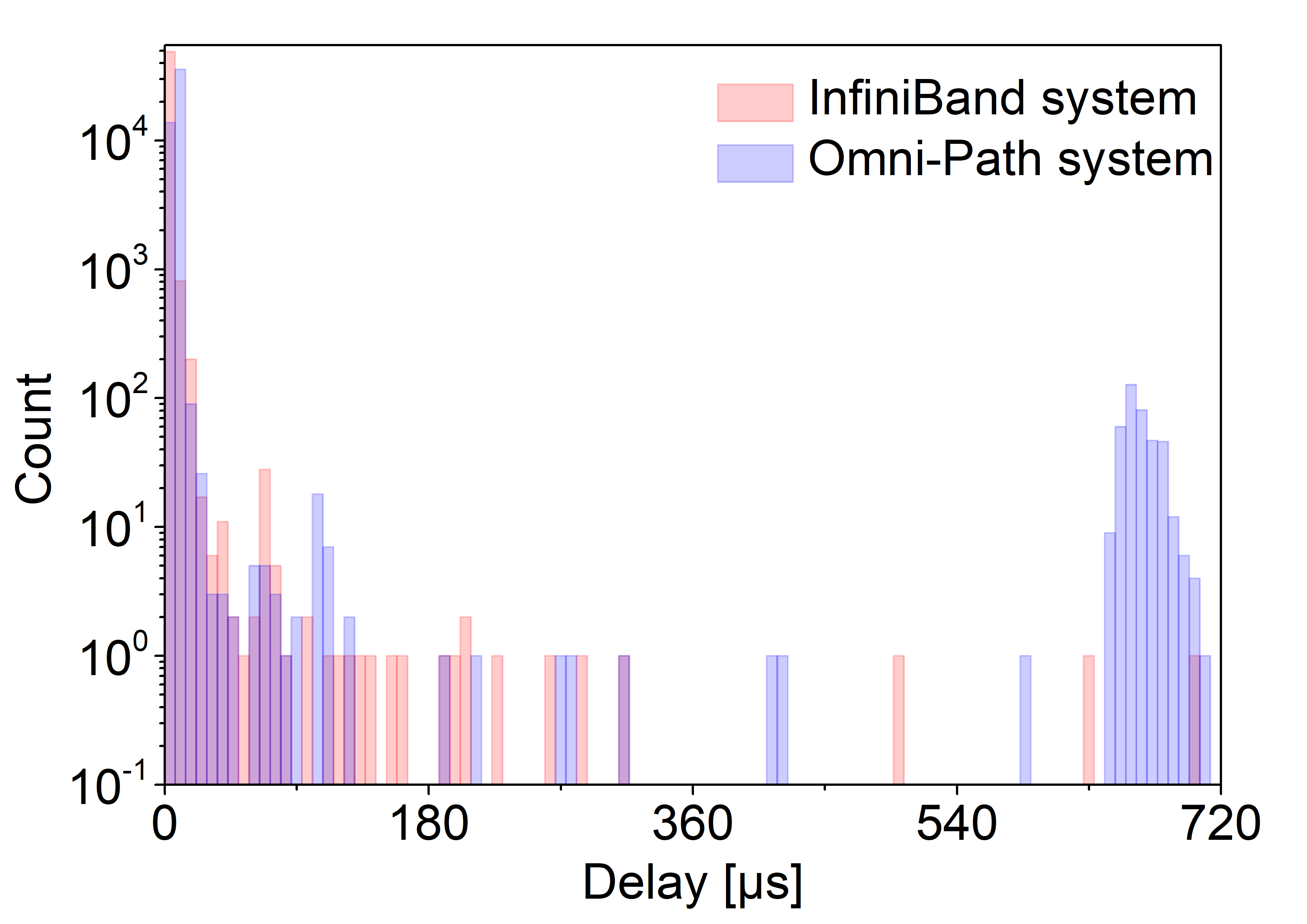}
	\caption{System noise with disabled SMT}
	\end{subfigure}
	\caption{Histograms for statistical baseline
          (natural/realistic) system-induced execution-level delays on
          InfiniBand (Emmy) and Omni-Path (Meggie) Systems over a
          period of 3\,ms, with and without \ac{SMT}. The bin size was
          640\,\ns\ (top) and 7.2\,\MuS\ (bottom), respectively.}
	\label{fig:Systemnoise}
\end{figure}
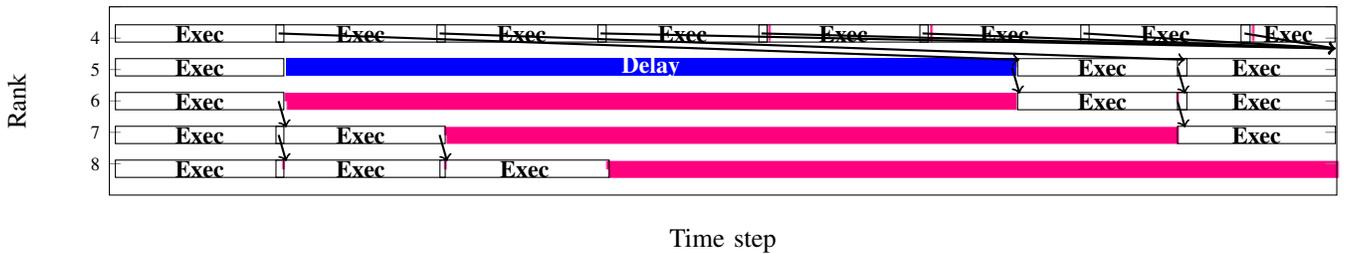
\begin{figure*}
	%	\begin{subfigure}[t]{0.375 \textheight} %height
	\begin{minipage}{\textwidth}
			\begin{tikzpicture}
			\put(6,2.8) {\includegraphics[width=.9\textwidth,height=0.098 \textheight]{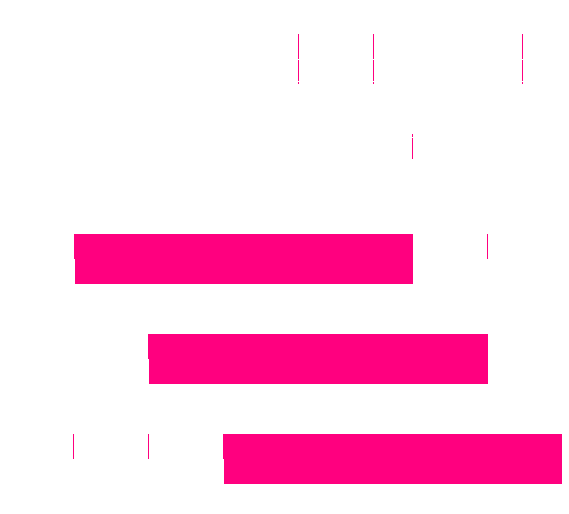}}
			\begin{axis}[
			width=.987\textwidth,height=0.173 \textheight,
			ylabel = {Rank},
			y label style={at={(0,0.5)}},
			xlabel = {Time step},
			x label style={at={(0.5,0.08)}},
			x tick label style={font=\scriptsize},
			y tick label style={font=\scriptsize},
			xmajorticks=false,
			xmin=1, xmax=8,
			ymin=0, ymax=6, 
			xtick={1,2,3,4,5,6,7,8,9,10,11,12,13,14,15,16,17,18,19,20,21,22,23,24,25},
			ytick={1,2,3,4,5,6,7},
			yticklabels={8,7,6,5,4},
			]
			\end{axis}
			\node [font=\small] at (7.2,1.7){\tikz \fill [blue] (10.2,0.4) rectangle (0.5,0.17) node[pos=.5, white] {\textbf{Delay}};}; 
			 \node [font=\small] at (1.2,2.15){\tikz \draw (2.7,0.4) rectangle (0.46,0.17) node[pos=.5] {\textbf{Exec}};};
			 \node [font=\small] at (1.2,1.7){\tikz \draw (2.7,0.4) rectangle (0.462,0.17) node[pos=.5] {\textbf{Exec}};};
			 \node [font=\small] at (1.2,1.25){\tikz \draw (2.7,0.4) rectangle (0.462,0.17) node[pos=.5] {\textbf{Exec}};}; 
			 \node [font=\small] at (1.2,0.8){\tikz \draw (2.7,0.4) rectangle (0.46,0.17) node[pos=.5] {\textbf{Exec}};}; 
			 \node [font=\small] at (1.2,0.35){\tikz \draw (2.7,0.4) rectangle (0.46,0.17) node[pos=.5] {\textbf{Exec}};};  
			 \node [font=\small] at (3.34,2.15){\tikz \draw (2.7,0.4) rectangle (0.45,0.17) node[pos=.5] {\textbf{Exec}};}; 
			 \node [font=\small] at (3.34,0.8){\tikz \draw (2.7,0.4) rectangle (0.45,0.17) node[pos=.5] {\textbf{Exec}};}; 
			 \node [font=\small] at (3.34,0.35){\tikz \draw (2.7,0.4) rectangle (0.45,0.17) node[pos=.5] {\textbf{Exec}};}; 
			 \node [font=\small] at (5.48,2.15){\tikz \draw (2.7,0.4) rectangle (0.45,0.17) node[pos=.5] {\textbf{Exec}};};
			 \node [font=\small] at (5.52,0.35){\tikz \draw (2.7,0.4) rectangle (0.45,0.17) node[pos=.5] {\textbf{Exec}};}; 
			 \node [font=\small] at (7.62,2.15){\tikz \draw (2.7,0.4) rectangle (0.45,0.17) node[pos=.5] {\textbf{Exec}};};
			\node [font=\small] at (13.2,1.7){\tikz \draw (2.7,0.4) rectangle (0.45,0.17) node[pos=.5] {\textbf{Exec}};};
			\node [font=\small] at (13.2,1.25){\tikz \draw (2.7,0.4) rectangle (0.45,0.17) node[pos=.5] {\textbf{Exec}};};
			 \node [font=\small] at (9.76,2.15){\tikz \draw (2.7,0.4) rectangle (0.45,0.17) node[pos=.5] {\textbf{Exec}};};
			\node [font=\small] at (15.25,1.7){\tikz \draw (2.55,0.4) rectangle (0.45,0.17) node[pos=.5] {\textbf{Exec}};};
			\node [font=\small] at (15.25,1.25){\tikz \draw (2.55,0.4) rectangle (0.45,0.17) node[pos=.5] {\textbf{Exec}};}; 
			\node [font=\small] at (15.25,0.8){\tikz \draw (2.55,0.4) rectangle (0.45,0.17) node[pos=.5] {\textbf{Exec}};};
			\node [font=\small] at (11.9,2.15){\tikz \draw (2.7,0.4) rectangle (0.45,0.17) node[pos=.5] {\textbf{Exec}};}; 
			 \node [font=\small] at (14.04,2.15){\tikz \draw (2.7,0.4) rectangle (0.45,0.17) node[pos=.5] {\textbf{Exec}};};
			 \node [font=\small] at (15.67,2.15){\tikz \draw (1.7,0.4) rectangle (0.45,0.17) node[pos=.5] {\textbf{Exec}};};
			\draw [thick, ->] (2.25,2.15) -- (12.1,1.8); 
			\draw [thick, ->] (2.25,1.25) -- (2.35,0.9); 
			\draw [thick, ->] (2.25,0.8) -- (2.35,0.45); 
			\draw [thick, ->] (4.39,2.15) -- (14.3,1.8); 
			\draw [thick, ->] (4.39,0.8) -- (4.49,0.45);
			\draw [thick, ->] (6.53,2.15) -- (16.3,1.95); 				
			\draw [thick, ->] (8.67,2.15) -- (16.3,1.95); 	
			\draw [thick, ->] (10.81,2.15) -- (16.3,1.95); 		
			\draw [thick, ->] (12.95,2.15) -- (16.3,1.95); 							
			\draw [thick, ->] (15.09,2.15) -- (16.3,1.95);
			\draw [thick, ->] (14.2,1.7) -- (14.3,1.35); 
			\draw [thick, ->] (14.2,1.25) -- (14.3,0.9);  
			\draw [thick, ->] (12,1.7) -- (12.1,1.35); 	
%			\draw [thick, ->] (12,1.25) -- (14,0.9);  
%			\draw [thick, ->] (14.2,0.8) -- (16.2,0.4); 											
%			\draw [dashed] (1.1,0) -- (0.6,-0.7);
			\end{tikzpicture}
		        \caption{The delay propagation mechanism in the most
                          simple setting: A long execution delay (spanning
                          several execution phases) is injected at MPI rank $5$
                          and time step $1$. Communication is in eager mode, and
                          unidirectional from process $i$ to $i+1$ after each
                          execution phase. The injected delay causes a waiting
                          phase (red bar) at rank $6$ and, after another
                          execution phase, at rank $7$, etc..  The idle wave
                          propagates through the system at a fixed speed due to
                          the regularity of execution
                          phases. Note that the width of the communication
                          phases has been exaggerated for clarity; communication
                          accounts for only about 0.2\% of the runtime.\label{fig:EagerUDP}}
	\end{minipage}	
\end{figure*}
The ``Emmy'' system is installed at Erlangen Regional Computing Center
(\ac{RRZE}).  It comprises $560$ dual-socket compute nodes with
ten-core Intel Xeon ``Ivy Bridge'' E5-2660v2 CPUs running at
\SI{2.2}{\giga \Hz}.  The fat-tree QDR InfiniBand interconnect
fabric (\SI{40}{\giga \bit \per \second} per link and direction) is
built on a hierarchy of 36-port switches.
``Meggie,'' also installed at \ac{RRZE}, features 724 dual-socket Intel Xeon ``Broadwell''
E5-2630v4 CPUs with ten cores each and a fat-tree Omni-Path network
(\SI{100}{\giga \bit \per \second} per link and direction).
For all measurements presented here, the clock frequency of all
nodes was fixed to the base value of 2.2\,\GHZ. Multi-node experiments
were run on a homogeneous set of nodes connected to a single
leaf switch. Process-core affinity was enforced using the available
facilities in the MPI implementation, ignoring the \ac{SMT} feature
(i.e., using only physical cores) unless specified otherwise.

We used the Intel \CPP{} compiler version $2019.2.187$ and the Intel
\ac{MPI} library version $2019$ (update $2$) for all experiments.
MPI traces were collected with the Intel Trace Analyzer and Collector
(same version). We
take advantage of \CPP{} high-resolution chrono clock for timing
measurement.
% and \ac{ITAC} for timeline visualization.
Communication delays for non-blocking calls were measured by time
spent in the \verb.MPI_Wait. wait function.

\subsection{System noise characteristics}

In order to know which levels and characteristics of noise are
realistic, an analysis of the natural system
noise on a standard compute node of each system was conducted. We ran an
MPI-parallel code with compute-bound workload whose execution time
(excluding communication) is known exactly. The code comprises a large number of back-to-back
double-precision divide instructions (\verb.vdivpd.), the throughput of which
is exactly one instruction per 28 clock cycles
on Ivy Bridge and one instruction per 16 clock cycles on Broadwell~\cite{Hofmann:2017}.
For the experiment we ran this workload on all physical cores of a node,
with execution phases of 3\,ms alternating with latency-bound
next-neighbor MPI communication, and measuring on each core
and for each execution phase the deviation of the pure execution time
from the ideal (ignoring the communication).
Overall, $3.3\times 10^5$ data points were collected.

Figure~\ref{fig:Systemnoise}(a) shows histograms of execution delays
with \ac{SMT} switched on. Both systems are quite similar, with average
delays of $2.4\,\MuS$ and $2.8\,\MuS$, respectively, and maximum delays
of less than $30\,\MuS$. With \ac{SMT} deactivated, however,
the Omni-Path system exhibits a bimodal distribution of noise
with a distictive second peak at $\approx 660\,\MuS$. We
speculate that this is the influence of the Omni-Path driver, which is much
more CPU-intensive than the InfiniBand driver on the other
system.  The damping
effect of \ac{SMT} on system noise is well known~\cite{Leon:2016}.
In our experiments we will use the clusters in their official
configuration, i.e., the InfiniBand cluster with SMT enabled
and the Omni-Path cluster with SMT disabled.

Note that system noise is an \emph{extrinsic} source of variation
from the application's point of view. There may also be
\emph{instrinsic} sources of noise such as random load imbalance
or variations in data access times. In later experiments we will inject
intrinsic fine-grained noise as part of the application execution,
but there is really no difference to extrinsic noise as far
as the observable outcome is concerned.

\section{Mechanisms of delay propagation}\label{sec:Mechanics}

\begin{figure*}[bt]
	\centering
	\begin{minipage}{\textwidth}
		\begin{subfigure}[t]{0.24\textwidth} %height
			\begin{tikzpicture}
			\put(-4.9,2.8) {\includegraphics[width=0.71\textwidth,height=0.098 \textheight]{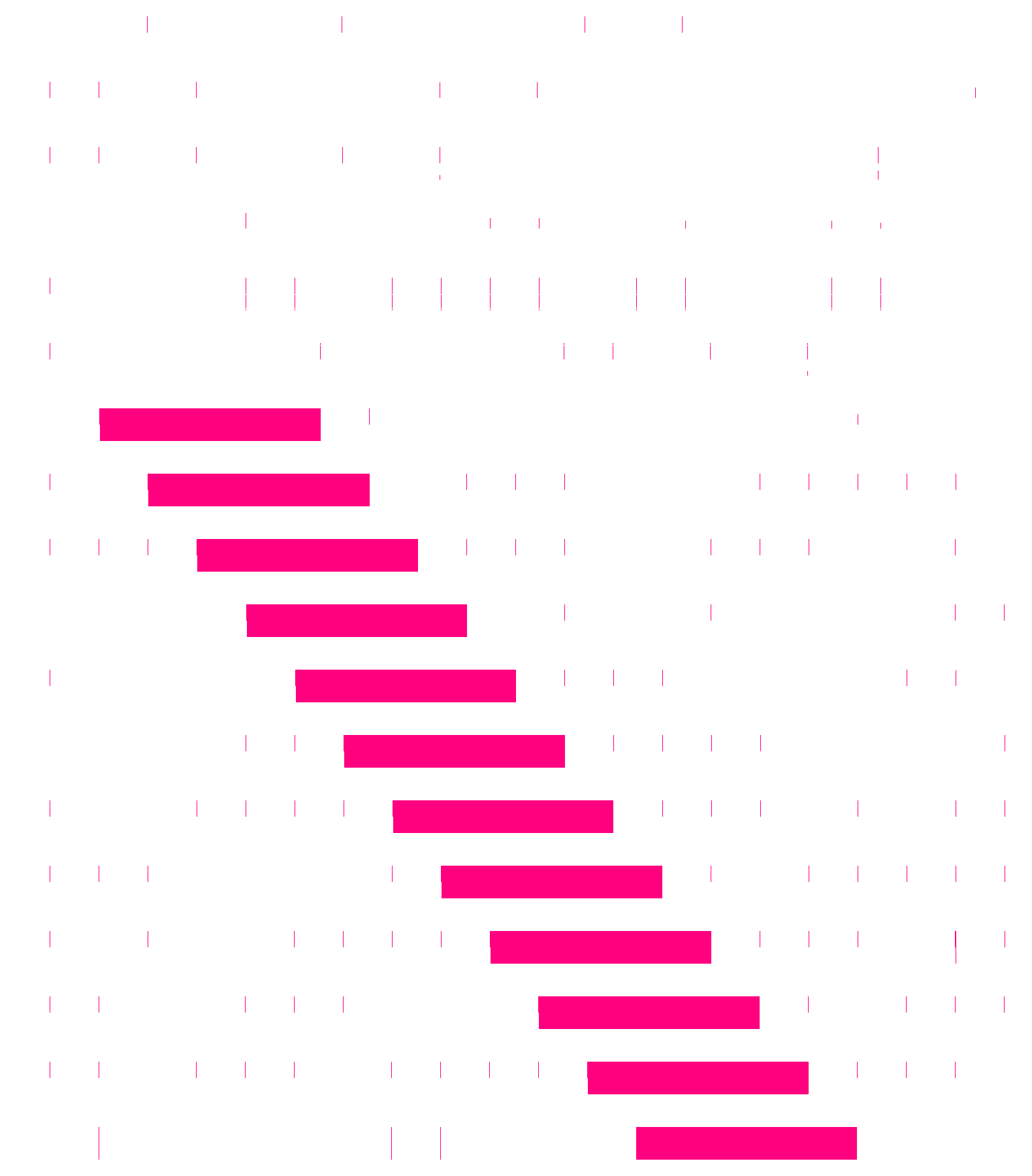}}
			\begin{axis}[
			width=1.1\textwidth,height=0.173 \textheight,
			ylabel = {Rank},
			y label style={at={(0.17,0.5)}},
			xlabel = {Time step},
			x label style={at={(0.5,0.08)}},
			x tick label style={font=\scriptsize},
			y tick label style={font=\scriptsize},
			xmin=1, xmax=25,
			ymin=0, ymax=19,
			xtick={1,2,3,4,5,6,7,8,9,10,11,12,13,14,15,16,17,18,19,20,21,22,23,24,25},
			ytick={1,2,3,4,5,6,7,8,9,10,11,12,13,14,15,16,17,18,19},
			xticklabels={,,2,,4,,6,,8,,10,,12,,,,\textbf{12},,14,,16,,18,,20},
			yticklabels={17,,15,,13,,11,,9,,7,,5,,3,,1,},
			name=DPM,
			]
%			\coordinate (x) at (axis cs:0.4,14.5);
%			\draw (x) rectangle (axis cs:10,9.5);
			\end{axis}
			\node [font=\small] at (1.5,-1.1){(a) Unidirectional open boundary}; 
			\node [font=\small] at (1.65,-1.5){$P_\mathrm{i}$ send to $P_\mathrm{i+1}$}; 
			\node [font=\small] at (1.7,-1.9){$P_\mathrm{i}$ receive from $P_\mathrm{i-1}$}; 
			\node [font=\small] at (0.45,1.71){\tikz \fill [blue] (1.66,0.1) rectangle (1,0.17);}; 
			\end{tikzpicture}
%			\put(0.1,7) {\tikz \fill [blue] (1.5,0.1) rectangle (1,0.2);}
		\end{subfigure}
		\begin{subfigure}[t]{0.24\textwidth} %height
			\begin{tikzpicture}
			\put(-4.8,2.8) {\includegraphics[width=0.875\textwidth,height=0.098 \textheight]{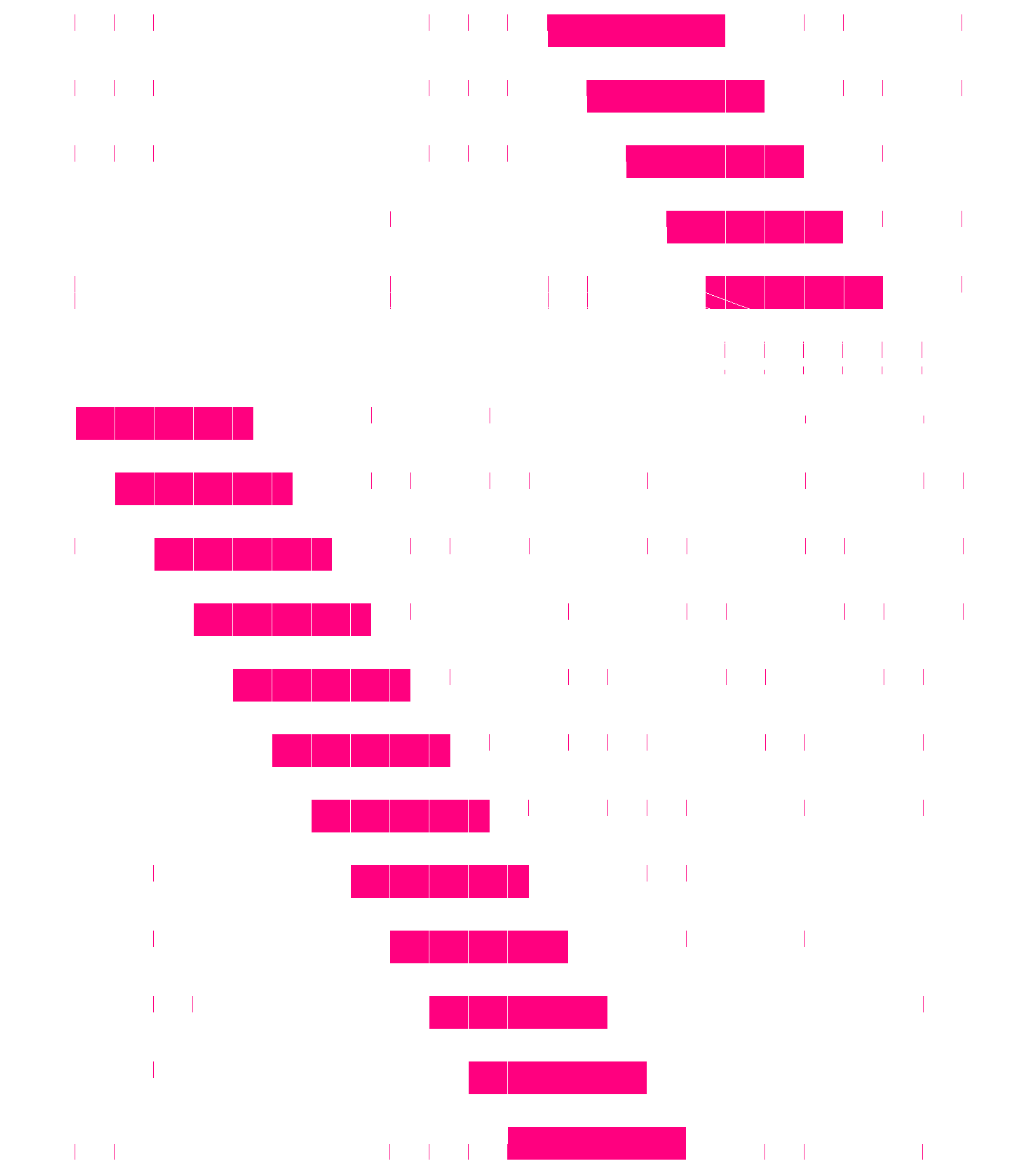}}
			\begin{axis}[
			width=1.188\textwidth,height=0.173 \textheight,
			ylabel = {Rank},
			y label style={at={(0.17,0.5)}},
			xlabel = {Time step},
			x label style={at={(0.5,0.08)}},
			x tick label style={font=\scriptsize},
			y tick label style={font=\scriptsize},
			xmin=1, xmax=25,
			ymin=0, ymax=19,
			xtick={1,2,3,4,5,6,7,8,9,10,11,12,13,14,15,16,17,18,19,20,21,22,23,24,25},
			ytick={1,2,3,4,5,6,7,8,9,10,11,12,13,14,15,16,17,18,19},
			xticklabels={,,2,,4,,6,,8,,10,,12,,,,12,,14,,,\textbf{17},,,20},
			yticklabels={17,,15,,13,,11,,9,,7,,5,,3,,1,},
			]
			\end{axis}
			\node [font=\small] at (1.5,-1.1){(b) Unidirectional periodic}; 
			\node [font=\small] at (1.65,-1.5){$P_\mathrm{i}$ send to $P_\mathrm{i+1}$}; 
			\node [font=\small] at (1.7,-1.9){$P_\mathrm{i}$ receive from $P_\mathrm{i-1}$}; 
			\node [font=\small] at (0.45,1.71){\tikz \fill [blue] (1.66,0.1) rectangle (1,0.17);}; 
			\end{tikzpicture}
		\end{subfigure}
		\begin{subfigure}[t]{0.24\textwidth} %height
			\begin{tikzpicture}
			\put(-4.8,2.8) {\includegraphics[width=0.875\textwidth,height=0.098 \textheight]{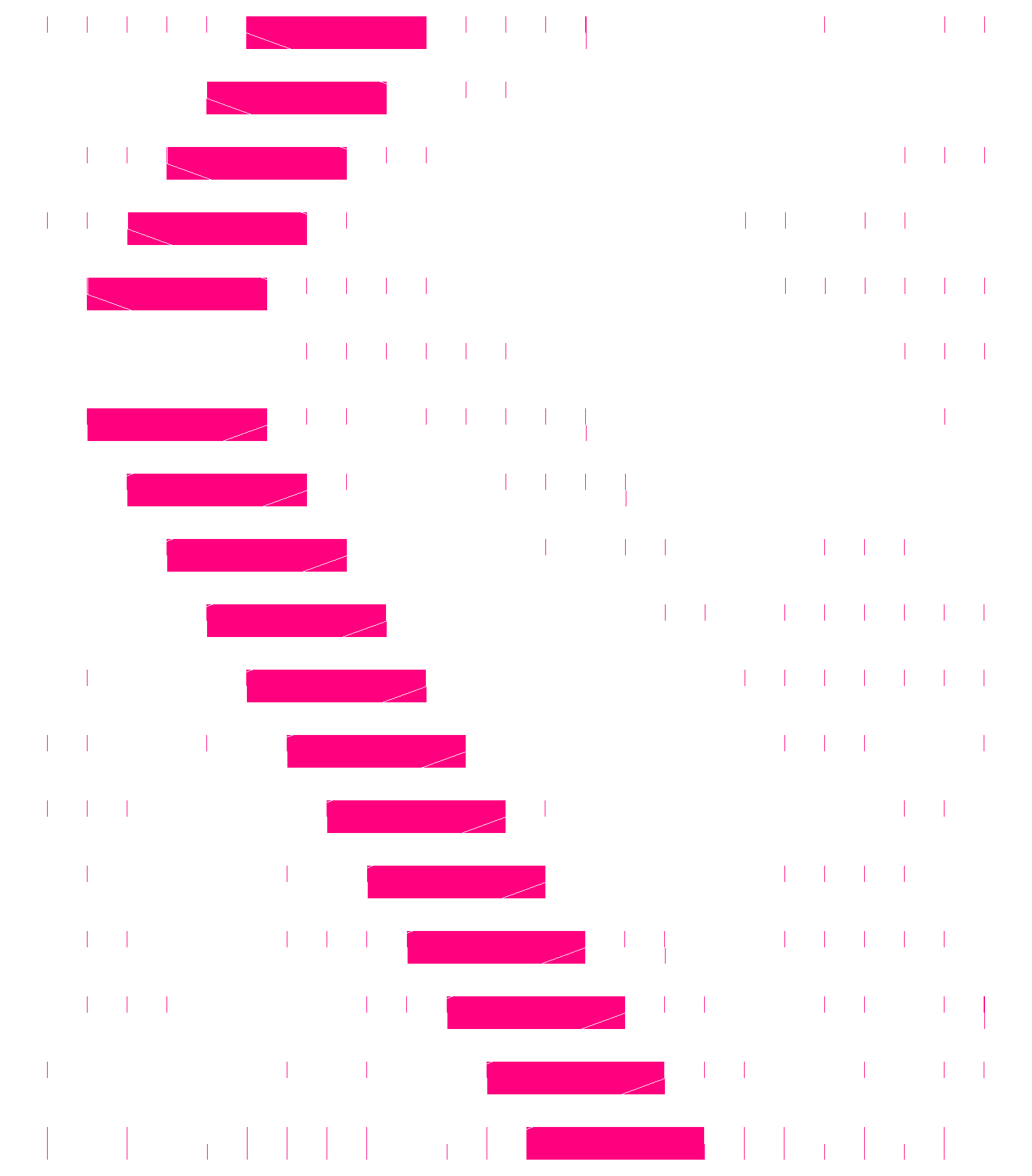}}
			\begin{axis}[
			width=1.188\textwidth,height=0.173 \textheight,
			ylabel = {Rank},
			y label style={at={(0.17,0.5)}},
			xlabel = {Time step},
			x label style={at={(0.5,0.08)}},
			x tick label style={font=\scriptsize},
			y tick label style={font=\scriptsize},
			xmin=1, xmax=25,
			ymin=0, ymax=19,
			xtick={1,2,3,4,5,6,7,8,9,10,11,12,13,14,15,16,17,18,19,20,21,22,23,24,25},
			ytick={1,2,3,4,5,6,7,8,9,10,11,12,13,14,15,16,17,18,19},
			xticklabels={,,2,,4,,6,,8,,10,,12,,,,\textbf{12},,14,,16,,18,,20},
			yticklabels={17,,15,,13,,11,,9,,7,,5,,3,,1,},
			]
			\end{axis}
			\node [font=\small] at (1.5,-1.1){(c) Bidirectional open boundary};
			\node [font=\small] at (1.65,-1.5){$P_\mathrm{i}$ send to $P_\mathrm{i\pm1}$}; 
			\node [font=\small] at (1.7,-1.9){$P_\mathrm{i}$ receive from $P_\mathrm{i\pm1}$}; 
			\node [font=\small] at (0.47,1.71){\tikz \fill [blue] (1.66,0.1) rectangle (1,0.17);}; 
			\end{tikzpicture}
		\end{subfigure}
		\begin{subfigure}[t]{0.24\textwidth} %height
			\begin{tikzpicture}
			\put(-4,2.8) {\includegraphics[width=0.875\textwidth,height=0.098 \textheight]{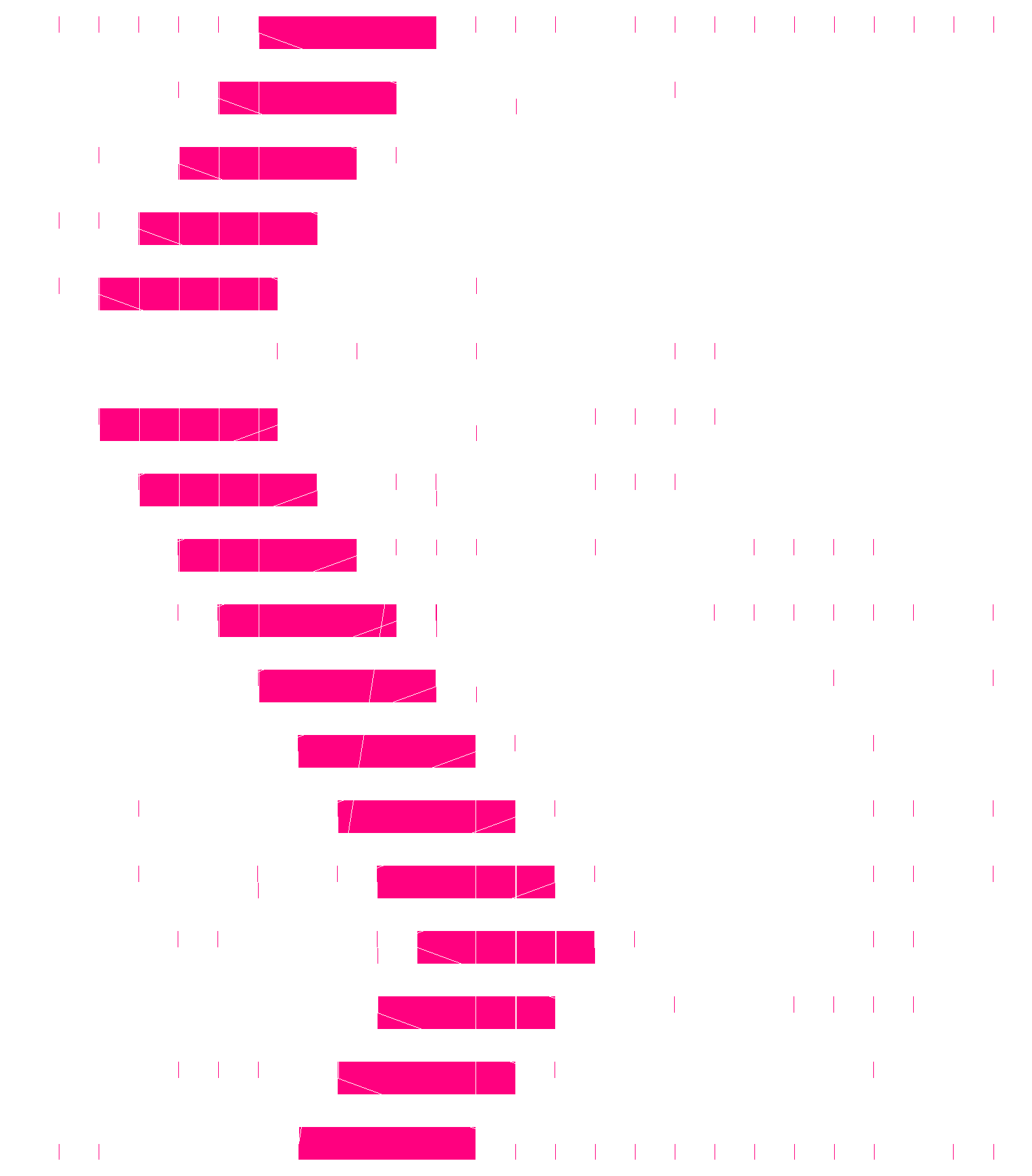}}
			\begin{axis}[
			width=1.188\textwidth,height=0.173 \textheight,
			ylabel = {Rank},
			y label style={at={(0.17,0.5)}},
			xlabel = {Time step},
			x label style={at={(0.5,0.08)}},
			x tick label style={font=\scriptsize},
			y tick label style={font=\scriptsize},
			xmin=1, xmax=25,
			ymin=0, ymax=19,
			xtick={1,2,3,4,5,6,7,8,9,10,11,12,13,14,15,16,17,18,19,20,21,22,23,24,25},
			ytick={1,2,3,4,5,6,7,8,9,10,11,12,13,14,15,16,17,18,19},
			xticklabels={,,2,,4,,6,,,,6,,,\textbf{9},,,12,,14,,16,,18,,20},
			yticklabels={17,,15,,13,,11,,9,,7,,5,,3,,1,},
			]
			\end{axis}
			\node [font=\small] at (1.5,-1.1){(d) Bidirectional periodic};
			\node [font=\small] at (1.65,-1.5){$P_\mathrm{i}$ send to $P_\mathrm{i\pm1}$}; 
			\node [font=\small] at (1.7,-1.9){$P_\mathrm{i}$ receive from $P_\mathrm{i\pm1}$};   
			\node [font=\small] at (0.53,1.71){\tikz \fill [blue] (1.66,0.1) rectangle (1,0.17);}; 
			\end{tikzpicture}
		\end{subfigure}
		\caption*{Small messages communication with eager protocol}
		%		\label{fig:Interaction}
		%\end{subfigure}
		%	\begin{subfigure}[t]{0.375 \textheight} 
	\end{minipage}
	\begin{minipage}{\textwidth}
		\begin{subfigure}[t]{0.24\textwidth} %height
			\begin{tikzpicture}
			\put(-6,3.5) {\includegraphics[width=0.875\textwidth,height=0.098 \textheight]{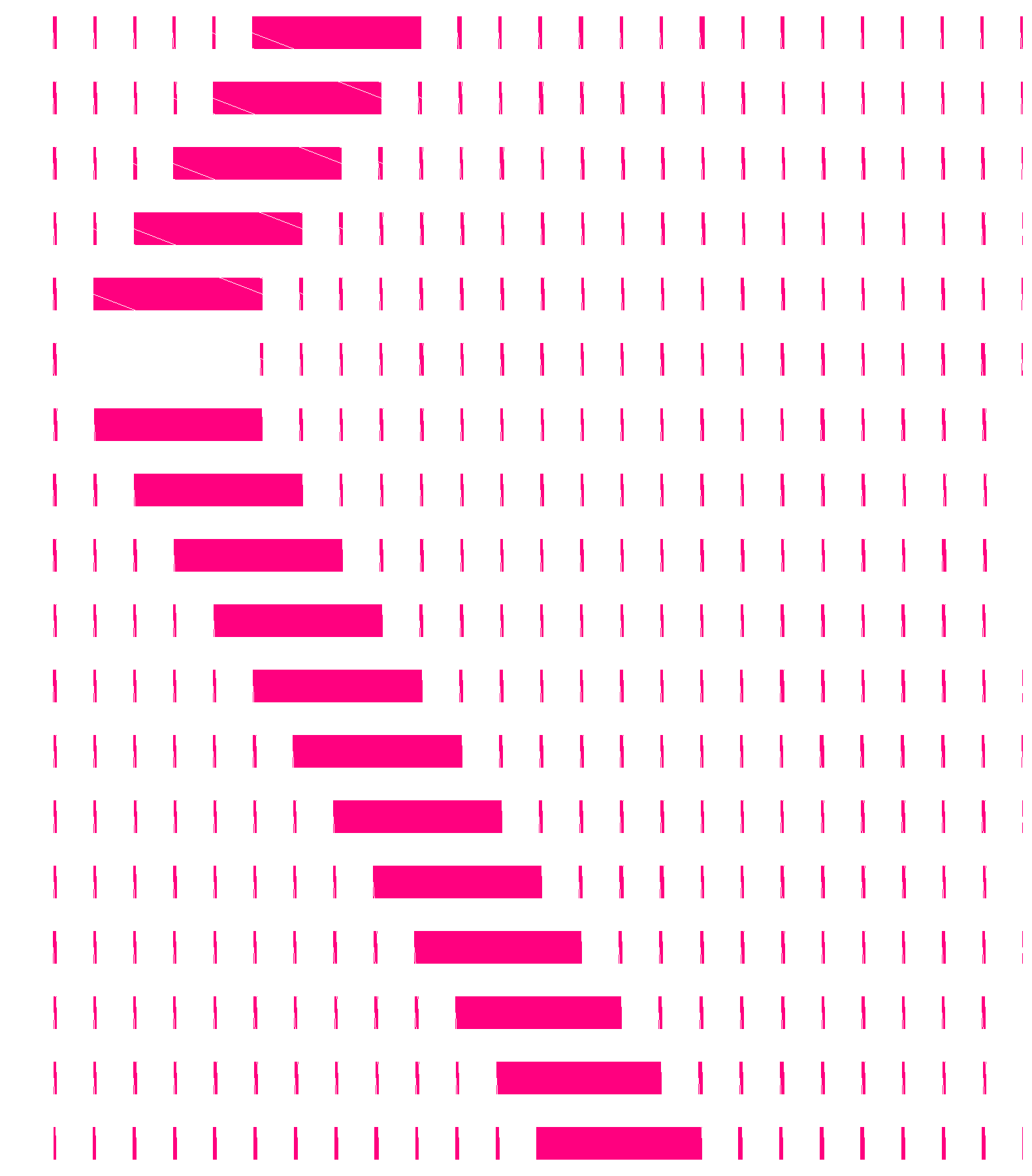}}
			\begin{axis}[
			width=1.188\textwidth,height=0.173 \textheight,
			ylabel = {Rank},
			y label style={at={(0.17,0.5)}},
			xlabel = {Time step},
			x label style={at={(0.5,0.08)}},
			x tick label style={font=\scriptsize},
			y tick label style={font=\scriptsize},
			xmin=1, xmax=25,
			ymin=0, ymax=19,
			xtick={1,2,3,4,5,6,7,8,9,10,11,12,13,14,15,16,17,18,19,20,21,22,23,24,25},
			ytick={1,2,3,4,5,6,7,8,9,10,11,12,13,14,15,16,17,18,19},
			xticklabels={,,2,,4,,6,,8,,10,,12,,,,\textbf{12},,14,,16,,18,,20},
			yticklabels={17,,15,,13,,11,,9,,7,,5,,3,,1,},
			]
			\end{axis}
			\node [font=\small] at (1.5,-1.1){(e) Unidirectional open boundary}; 
			\node [font=\small] at (1.65,-1.5){$P_\mathrm{i}$ send to $P_\mathrm{i+1}$}; 
			\node [font=\small] at (1.7,-1.9){$P_\mathrm{i}$ receive from $P_\mathrm{i-1}$}; 
			\node [font=\small] at (0.45,1.74){\tikz \fill [blue] (1.63,0.1) rectangle (1,0.17);}; 
			\end{tikzpicture}
		\end{subfigure}
		\begin{subfigure}[t]{0.24\textwidth} %height
			\begin{tikzpicture}
			\put(-5,3.5) {\includegraphics[width=0.875\textwidth,height=0.098 \textheight]{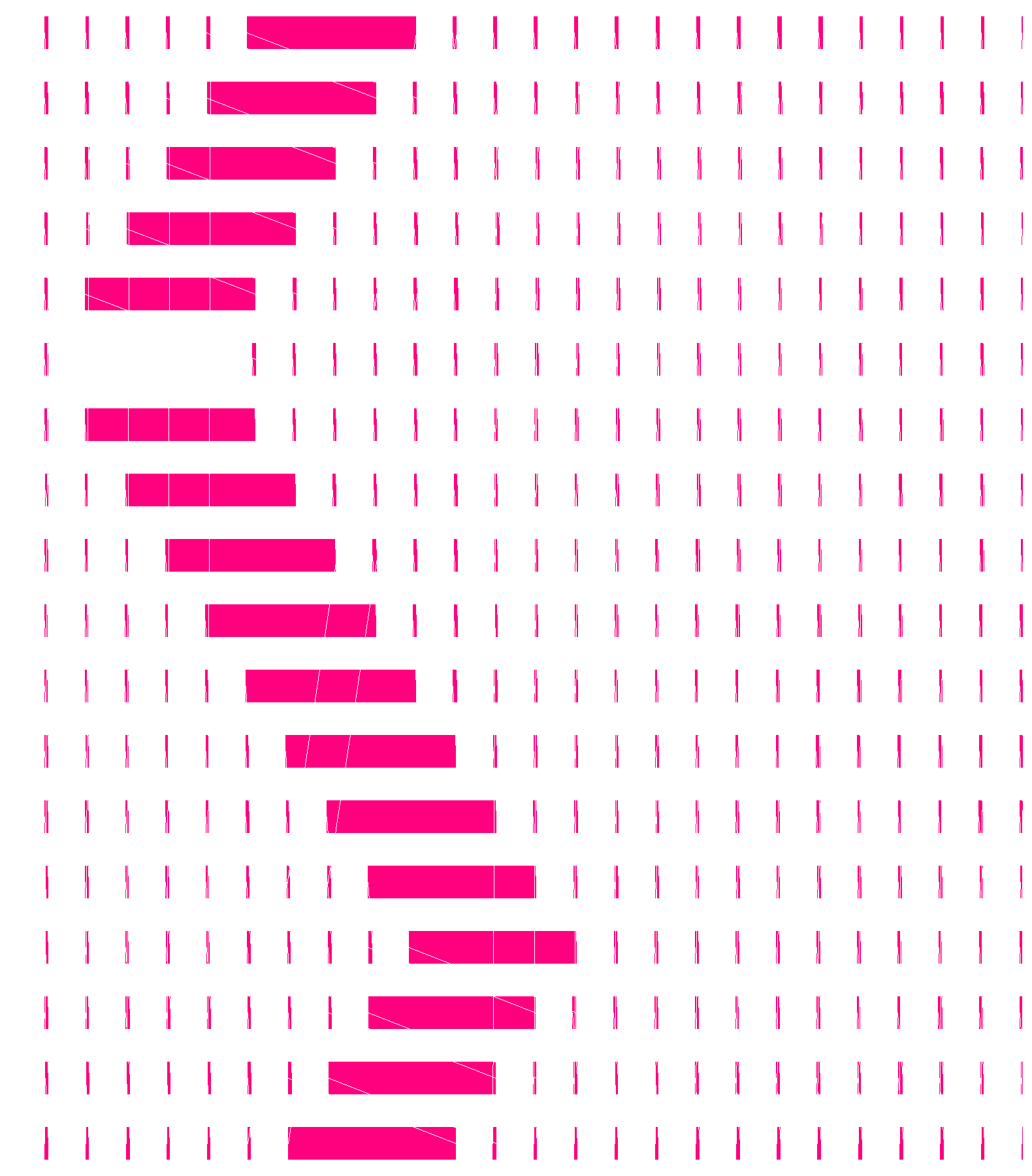}}
			\begin{axis}[
			width=1.188\textwidth,height=0.173 \textheight,
			ylabel = {Rank},
			y label style={at={(0.17,0.5)}},
			xlabel = {Time step},
			x label style={at={(0.5,0.08)}},
			x tick label style={font=\scriptsize},
			y tick label style={font=\scriptsize},
			xmin=1, xmax=25,
			ymin=0, ymax=19,
			xtick={1,2,3,4,5,6,7,8,9,10,11,12,13,14,15,16,17,18,19,20,21,22,23,24,25},
			ytick={1,2,3,4,5,6,7,8,9,10,11,12,13,14,15,16,17,18,19},
			xticklabels={,,2,,4,,6,,,,6,,,\textbf{9},,,12,,14,,16,,18,,20},
			yticklabels={17,,15,,13,,11,,9,,7,,5,,3,,1,},
			]
			\end{axis}
			\node [font=\small] at (1.5,-1.1){(f) Unidirectional periodic}; 
			\node [font=\small] at (1.65,-1.5){$P_\mathrm{i}$ send to $P_\mathrm{i+1}$}; 
			\node [font=\small] at (1.7,-1.9){$P_\mathrm{i}$ receive from $P_\mathrm{i-1}$}; 
			\node [font=\small] at (0.45,1.74){\tikz \fill [blue] (1.63,0.1) rectangle (1,0.17);}; 
			\end{tikzpicture}
		\end{subfigure}
		\begin{subfigure}[t]{0.24\textwidth} %height
			\begin{tikzpicture}
			\put(-5.6,3.5) {\includegraphics[width=0.875\textwidth,height=0.098 \textheight]{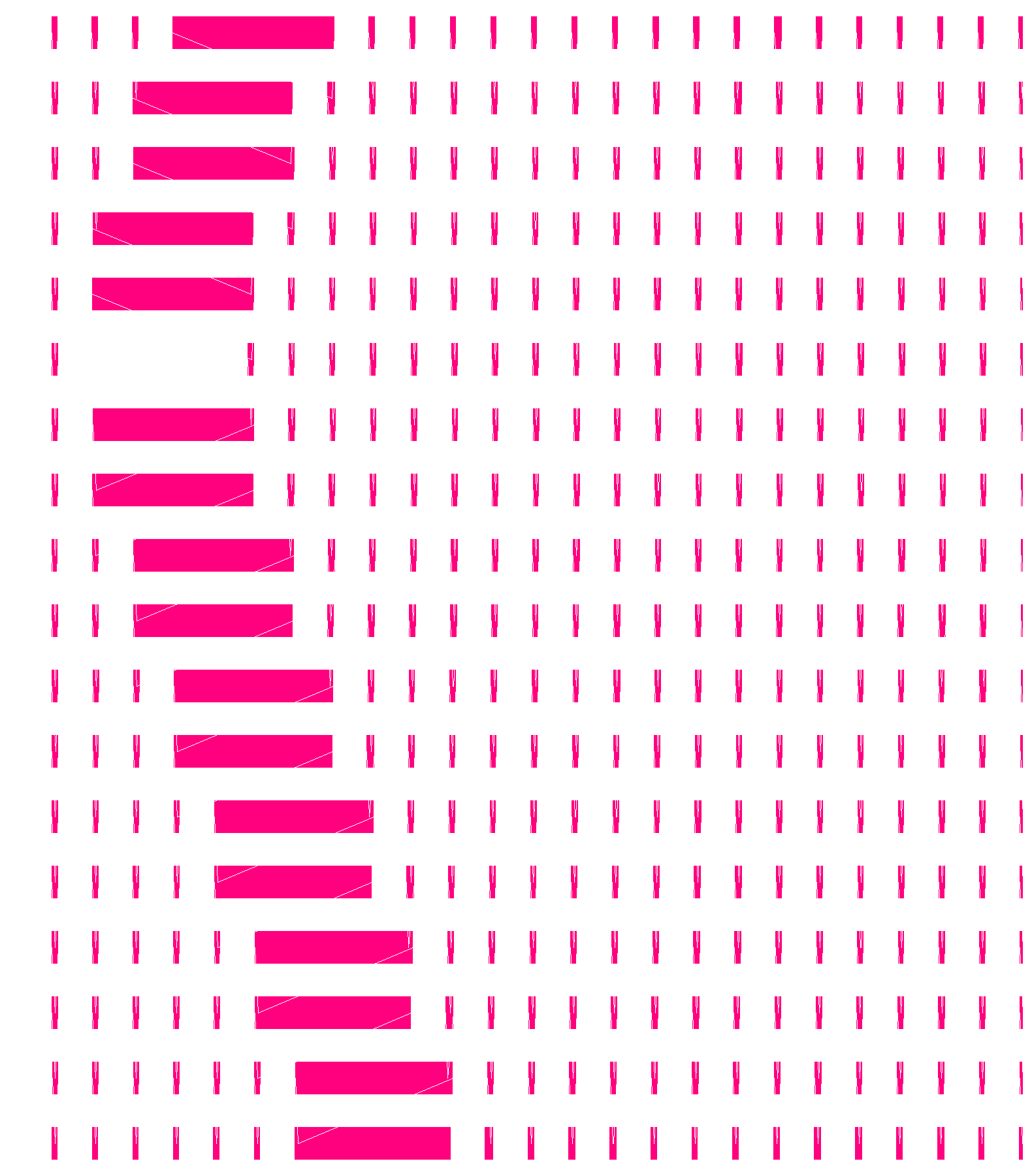}}
			\begin{axis}[
			width=1.188\textwidth,height=0.173 \textheight,
			ylabel = {Rank},
			y label style={at={(0.17,0.5)}},
			xlabel = {Time step},
			x label style={at={(0.5,0.08)}},
			x tick label style={font=\scriptsize},
			y tick label style={font=\scriptsize},
			xmin=1, xmax=25,
			ymin=0, ymax=19,
			xtick={1,2,3,4,5,6,7,8,9,10,11,12,13,14,15,16,17,18,19,20,21,22,23,24,25},
			ytick={1,2,3,4,5,6,7,8,9,10,11,12,13,14,15,16,17,18,19},
			xticklabels={,,2,,4,,6,,,,\textbf{6},,8,,10,,12,,14,,16,,18,,20},
			yticklabels={17,,15,,13,,11,,9,,7,,5,,3,,1,},
			]
			\end{axis}
			\node [font=\small] at (1.5,-1.1){(g) Bidirectional open boundary}; 
			\node [font=\small] at (1.65,-1.5){$P_\mathrm{i}$ send to $P_\mathrm{i\pm1}$}; 
			\node [font=\small] at (1.7,-1.9){$P_\mathrm{i}$ receive from $P_\mathrm{i\pm1}$}; 
			\node [font=\small] at (0.44,1.74){\tikz \fill [blue] (1.6,0.1) rectangle (1,0.17);}; 
			\end{tikzpicture}
		\end{subfigure}
		\begin{subfigure}[t]{0.24\textwidth} %height
			\begin{tikzpicture}
			\put(-4.8,3.5) {\includegraphics[width=0.875\textwidth,height=0.098 \textheight]{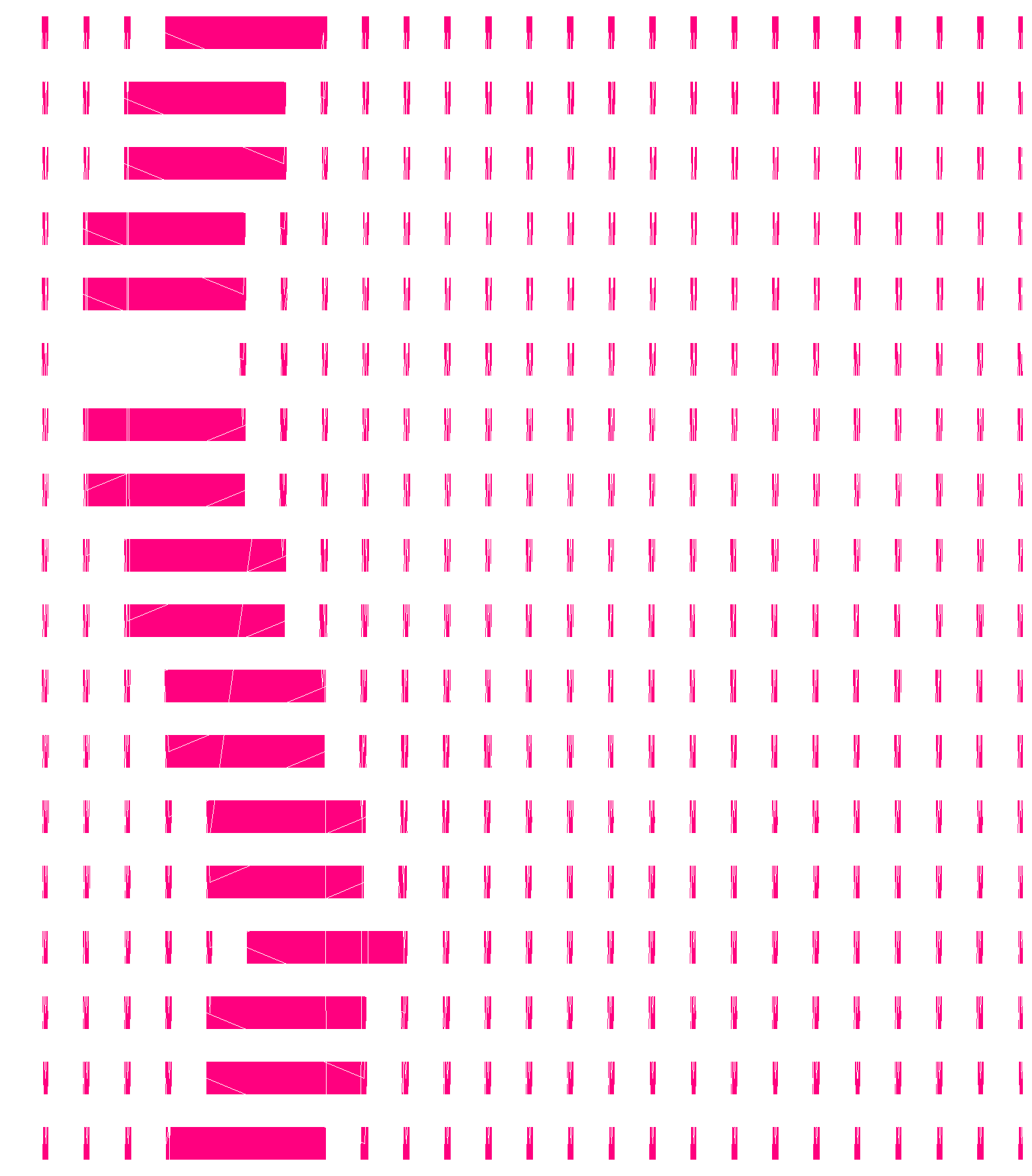}}
			\begin{axis}[
			width=1.188\textwidth,height=0.173 \textheight,
			ylabel = {Rank},
			y label style={at={(0.17,0.5)}},
			xlabel = {Time step},
			x label style={at={(0.5,0.08)}},
			x tick label style={font=\scriptsize},
			y tick label style={font=\scriptsize},
			xmin=1, xmax=25,
			ymin=0, ymax=19,
			xtick={1,2,3,4,5,6,7,8,9,10,11,12,13,14,15,16,17,18,19,20,21,22,23,24,25},
			ytick={1,2,3,4,5,6,7,8,9,10,11,12,13,14,15,16,17,18,19},
			xticklabels={,,,3,,,,3,,\textbf{5},,,8,,10,,12,,14,,16,,18,,20},
			yticklabels={17,,15,,13,,11,,9,,7,,5,,3,,1,},
			]
			\end{axis}
			\node [font=\small] at (1.5,-1.1){(h) Bidirectional periodic}; 
			\node [font=\small] at (1.65,-1.5){$P_\mathrm{i}$ send to $P_\mathrm{i\pm1}$}; 
			\node [font=\small] at (1.7,-1.9){$P_\mathrm{i}$ receive from $P_\mathrm{i\pm1}$}; 
			\node [font=\small] at (0.44,1.74){\tikz \fill [blue] (1.6,0.1) rectangle (1,0.17);}; 
			\end{tikzpicture}
		\end{subfigure}
		\caption*{Large messages communication with rendezvous protocol}
	\end{minipage}
	\caption{Qualitative timeline analysis of delay propagation under
          controlled conditions on the InfiniBand cluster with one process per
          node, next-neighbor non-blocking communication, and
          only native system noise.  Execution and communication phases are
          shown in white, blue is a deliberately injected delay at rank $5$, and
          idle periods and communication delays are red.  The message size for small
          messages (top row) was \SI{16384}{\byte} and for large messages
          (bottom row) \SI{31080}{\byte}, the eager limit being at
          \SI{16384}{doubles}, i.e., \SI{131072}{\byte}.
          For each message size, all four combinations of
          uni/bidirectional and periodic/open boundary conditions are shown.}
	\label{fig:homo}
\end{figure*}

We are especially interested in the propagation behavior of long execution
delays on the execution time of regular, bulk-synchronous applications and how
intrinsic or extrinsic fine-grained noise influences this propagation.  To this
end, we performed a series of experiments to fathom a part of the vast parameter
space. It turned out that the Ivy Bridge InfiniBand cluster (``Emmy'') showed
behavior that is almost perfectly in line with the
LogGOPSim~\cite{hoefler-loggopsim} simulator, so we use the ``real''
systems throughout unless indicated otherwise.
One execution phase is purely
compute bound and 3\,ms long, and the message size is 8192\,byte
if no other specification is given.
Figure~\ref{fig:EagerUDP} shows the most simple case:
eager-mode, unidirectional communication, one process per node,
no significant system or application
noise. Each rank $i$ sends a message to the next, $i+1$, and cannot
continue before it has received a message from $i-1$.
A delay of a length of 4.5 execution phases (blue bar) is injected at rank
5 at the first time step. Due to the delayed message from rank 5 to rank 6,
the latter gets delayed by the same amount, and so is the message it sends to
rank 7 at the end of the next execution phase, etc..
Due to the eager protocol, ranks smaller than 5 are unaffected by the delay
because they can get rid of their messages.\footnote{There is of course a
  limit to the internal buffers that store such messages, but this can be handled
  like a transition to a rendezvous protocol.}
In effect, the injected
delay causes an ``idle wave''~\cite{markidis2015idle} to ripple through the
system at a constant speed of one rank per execution plus communication phase length.
Note that this is only strictly true for homogeneous systems
and core-bound execution. The presence of a memory bottleneck and/or
different domains with different communication characteristics (e.g.,
intranode vs.\ internode) will change the picture, but this
is outside the scope of this work.

%%%%%%%%%%%%%%%%%%%%%%%%%%%%%%%%%%%%%%%%%%%%%%%%%%%%%%%%%%%%%%%%%%%%%%%%%%%%%%%

\subsection{Basic flavors of delay propagation}\label{sec:Flavors}

It can be expected that the different communication parameters described in
Section~\ref{sec:Communication} cause different idle wave propagation
patterns. Figure~\ref{fig:homo} shows a scan of all eight combinations of
eager/rendezvous protocol, periodic/open boundary conditions, and
unidirectional/bidirectional non-blocking next-neighbor communication, again running only one
process per node. In all cases the communication was implemented
by first initiating nonblocking \verb.MPI_Isend./\verb.MPI_Irecv.
calls to all neighbors of a process and then calling \verb.MPI_Waitall.\@. 

Figure~\ref{fig:homo}(a) depicts the same situation as in
Figure~\ref{fig:EagerUDP} (eager, unidirectional, nonperiodic)  but shown for the
full set of 18 ranks. Due to the
open boundary conditions, the idle wave runs out at the last process.
This changes with periodic boundaries (Figure~\ref{fig:homo}(b)): The
idle wave wraps around until, after 17 steps, it hits the process on which
the delay was injected. There it dies out because process 5 is still busy
receiving the outstanding eager messages from above, and as soon
as the idle period on rank 4 ends, everything is in sync again.

Figures~\ref{fig:homo}(c) and (d) show the situation for eager but bidirectional
communication. Idle waves must now propagate in both directions from the
injection but die out at the boundary for a nonperiodic process chain. In the
periodic case, they wrap around and meet at rank 14 where they cancel. This is
the first indication that idle waves must be a \emph{nonlinear} phenomenon that
cannot be adequately described by a linear wave equation.

With larger messages, the rendezvous protocol kicks in (lower row in
Figure~\ref{fig:homo}). Now even with unidirectional communication
(Figures~\ref{fig:homo}(e), (f)) the idle wave must propagate in both directions
because rank 4 cannot get rid of its messages to rank 5 as long as the injected
delay lasts. The general pattern is thus the same as for bidirectional
eager-mode communication (Figures~\ref{fig:homo}(c), (d)).

Finally, with bidirectional rendezvous-mode communication
(Figures~\ref{fig:homo}(g), (h)) the idle wave propagates
twice as fast, because two neighbors of the delayed process
are blocked in either direction.

These observations are entirely expected when looking at the
basic mechanisms of point-to-point communication, but
several questions come to mind: Does the interaction of
propagating idle periods have a more intricate phenomenology
than shown with these simple and controlled experiments?
What is the speed by which an idle wave ripples through the system?
Does system or application noise change the overall picture?
And what is the role of system topology, specifically
the intranode multicore structure of the cluster? These
questions will be addressed in the following sections.

%%%%%%%%%%%%%%%%%%%%%%%%%%%%%%%%%%%%%%%%%%%%%%%%%%%%%%%%%%%%%%%%%%%%%%%%%%%%%%%

\subsection{Interaction of propagating delays}

\begin{figure}[tb]
\begin{subfigure}[t]{0.15\textwidth} %height
	\begin{tikzpicture}
	\put(-2.9,0.42) {\includegraphics[width=0.86\textwidth,height=0.1905 \textheight]{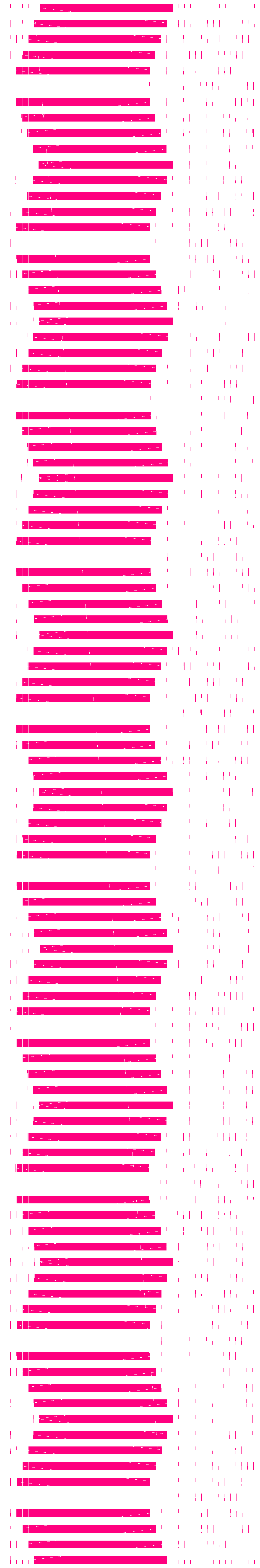}}
	\begin{axis}[
	width=1.4\textwidth,height=0.258\textheight,
	ylabel = {Rank},
	y label style={at={(0.17,0.5)}},
	xlabel = {Time step},
	x label style={at={(0.5,0.08)}},
	x tick label style={font=\scriptsize},
	y tick label style={font=\scriptsize}, 
	xmin=1, xmax=35,
	ymin=0, ymax=99,
	xtick={1,2,3,4,5,6,7,8,9,10,11,12,13,14,15,16,17,18,19,20,21,22,23,24,25,26,27,28,29,30,31,32,33,34,35},
	ytick={4,9,14,19,24,29,34,39,44,49,54,59,64,69,74,79,84,89,94}, 
	xticklabels={,,,,,,,,,,,,,,,,,,,,,,,,,,,,,,,,,20},
yticklabels={\textbf{95},,\textbf{85},,\textbf{75},,\textbf{65},,\textbf{55},,\textbf{45},,\textbf{35},,\textbf{25},,\textbf{15},,\textbf{5}},
	]
	\end{axis}
			\node at (1,-1.1){(a) equal}; 
			\draw [semithick, dotted] (-0.6,3.65) -- (2.2,3.65); 
			\draw [semithick, dotted] (-0.6,2.7375) -- (2.2,2.7375); 
			\draw [semithick, dotted] (-0.6,1.825) -- (2.2,1.825); 
			\draw [semithick, dotted] (-0.6,4.10625) -- (2.2,4.10625); 
			\draw [semithick, dotted] (-0.6,3.19375 ) -- (2.2,3.19375 ); 
			\draw [semithick, dotted] (-0.6,2.28125 ) -- (2.2,2.28125 );
			\draw [semithick, dotted] (-0.6,1.36875) -- (2.2,1.36875);      
			\draw [semithick, dotted] (-0.6,0.9125) -- (2.2,0.9125); 
			\draw [semithick, dotted] (-0.6,0.45625) -- (2.2,0.45625); 			
	\end{tikzpicture}
\end{subfigure}
\begin{subfigure}[t]{0.15\textwidth} %height
	\begin{tikzpicture}
	\put(-2.2,0.42) {\includegraphics[width=0.859\textwidth,height=0.1905 \textheight]{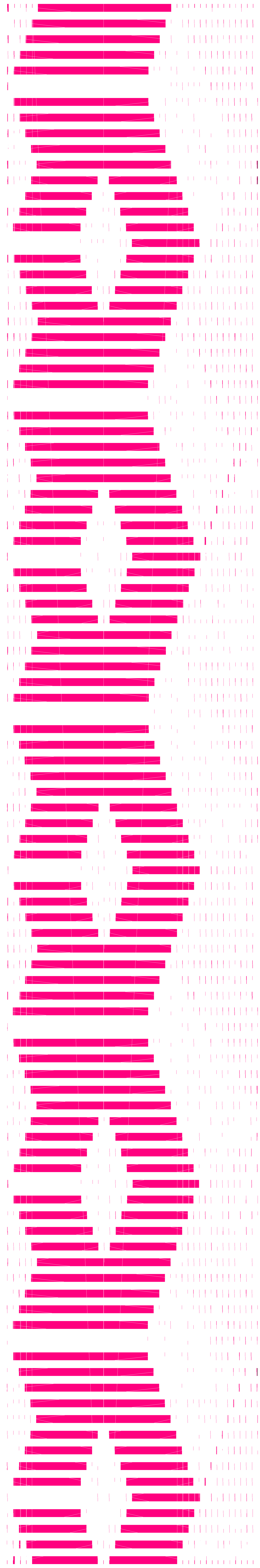}}
	\begin{axis}[
	width=1.4\textwidth,height=0.258\textheight,
	y label style={at={(0.17,0.5)}},
	xlabel = {Time step},
	x label style={at={(0.5,0.08)}},
	x tick label style={font=\scriptsize},
	y tick label style={font=\scriptsize},
	xmin=1, xmax=35,
	ymin=0, ymax=99,
	xtick={1,2,3,4,5,6,7,8,9,10,11,12,13,14,15,16,17,18,19,20,21,22,23,24,25,26,27,28,29,30,31,32,33,34,35},
	ytick={4,9,14,19,24,29,34,39,44,49,54,59,64,69,74,79,84,89,94}, 
xticklabels={,,,,,,,,,,,,,,,,,,,,,,,,,,,,,,,,,20},
yticklabels={\textbf{95},,\textbf{85},,\textbf{75},,\textbf{65},,\textbf{55},,\textbf{45},,\textbf{35},,\textbf{25},,\textbf{15},,\textbf{5}},
	]
	\end{axis}
	\node at (1,-1.1){(b) half};  
	\draw [semithick, dotted] (-0.6,3.65) -- (2.2,3.65); 
	\draw [semithick, dotted] (-0.6,2.7375) -- (2.2,2.7375); 
	\draw [semithick, dotted] (-0.6,1.825) -- (2.2,1.825); 
	\draw [semithick, dotted] (-0.6,4.10625) -- (2.2,4.10625); 
	\draw [semithick, dotted] (-0.6,3.19375 ) -- (2.2,3.19375 ); 
	\draw [semithick, dotted] (-0.6,2.28125 ) -- (2.2,2.28125 );
	\draw [semithick, dotted] (-0.6,1.36875) -- (2.2,1.36875);      
	\draw [semithick, dotted] (-0.6,0.9125) -- (2.2,0.9125); 
	\draw [semithick, dotted] (-0.6,0.45625) -- (2.2,0.45625); 	
%	\node [font=\small] at (0.35,1.12){\tikz \fill [blue] (1.63,0.1) rectangle (1,0.12);}; 
%\node [font=\small] at (0.35,0.22){\tikz \fill [blue] (1.63,0.1) rectangle (1,0.12);}; 
	\end{tikzpicture}
\end{subfigure}
\begin{subfigure}[t]{0.15\textwidth} %height 
	\begin{tikzpicture}
	\put(-1.7,0.42) {\includegraphics[width=0.84\textwidth,height=0.1905 \textheight]{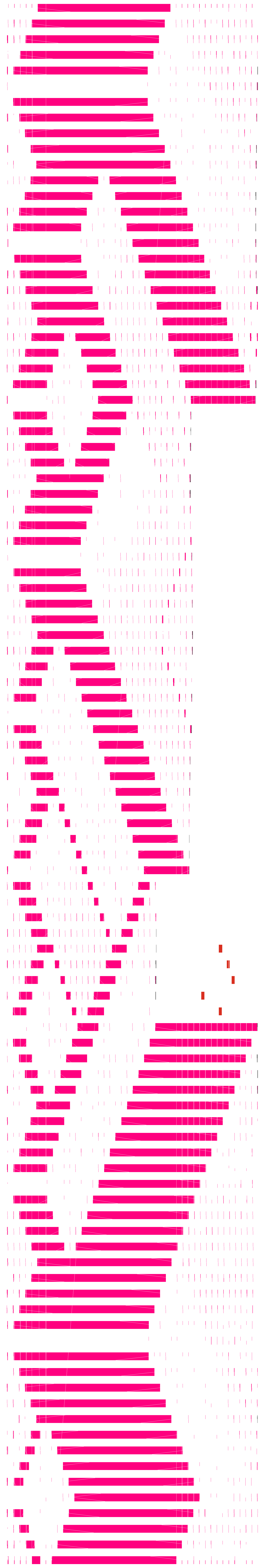}}
	\begin{axis}[
	width=1.4\textwidth,height=0.258\textheight,
	y label style={at={(0.17,0.5)}},
	xlabel = {Time step},
	x label style={at={(0.5,0.08)}},
	x tick label style={font=\scriptsize},
	y tick label style={font=\scriptsize},
	xmin=1, xmax=35,
	ymin=0, ymax=99,
	xtick={1,2,3,4,5,6,7,8,9,10,11,12,13,14,15,16,17,18,19,20,21,22,23,24,25,26,27,28,29,30,31,32,33,34,35},
	ytick={4,9,14,19,24,29,34,39,44,49,54,59,64,69,74,79,84,89,94}, 
	xticklabels={,,,,,,,,,,,,,,,,,,,,,,,,,,,,,,,,,20},
yticklabels={\textbf{95},,\textbf{85},,\textbf{75},,\textbf{65},,\textbf{55},,\textbf{45},,\textbf{35},,\textbf{25},,\textbf{15},,\textbf{5}},
	]
	\end{axis}
	\node at (1,-1.1){(c) random}; 
				\draw [semithick, dotted] (-0.6,3.65) -- (2.2,3.65); 
	\draw [semithick, dotted] (-0.6,2.7375) -- (2.2,2.7375); 
	\draw [semithick, dotted] (-0.6,1.825) -- (2.2,1.825); 
	\draw [semithick, dotted] (-0.6,4.10625) -- (2.2,4.10625); 
	\draw [semithick, dotted] (-0.6,3.19375 ) -- (2.2,3.19375 ); 
	\draw [semithick, dotted] (-0.6,2.28125 ) -- (2.2,2.28125 );
	\draw [semithick, dotted] (-0.6,1.36875) -- (2.2,1.36875);      
	\draw [semithick, dotted] (-0.6,0.9125) -- (2.2,0.9125); 
	\draw [semithick, dotted] (-0.6,0.45625) -- (2.2,0.45625); 	
	\end{tikzpicture}
\end{subfigure}
\caption{Qualitative analysis of interacting idle waves in a periodic
  process chain. Eager-mode bidirectional communication (message size
  \SI{16384}{\byte}) was used on the InfiniBand cluster, and ten
  processes per socket were run on 10 sockets (5 nodes).
  (a) Same delay injected at sixth process on each socket,
  (b) delay injected at sixth process on each socket, but
  delay duration was half on odd sockets,
  (c) random delay injected at sixth process of each socket.
  Dotted lines mark socket boundaries.}
\label{fig:hetro}
\end{figure}
As demonstrated in the previous section, idle waves can run out at process chain
boundaries or cancel completely when hitting each other.  Since delays of
different duration might be injected in random ways across the whole
communicator, the question arises what happens in more complex
scenarios. Figure~\ref{fig:hetro} shows the result of three experiments on
100 MPI processes with bidirectional eager-mode communication and periodic
boundary conditions running on ten sockets (five nodes) of the InfiniBand
cluster. Of course, the intra-node communication characteristics differ from the
InfiniBand parameters, but this is of no significance here.
Delays were injected on local rank 5 of every socket. For equal delays
(Figure~\ref{fig:hetro}(a)) we observe the expected cancellation after five
hops. If delays on odd sockets are just half as long
(Figure~\ref{fig:hetro}(b)), \emph{partial cancellation} occurs and the
(originally) longer idle periods continue to propagate until they cancel with
their symmetric counterparts from the next even socket. With random
injections (Figure~\ref{fig:hetro}(c)), the longest initial delays cause
idle waves that survive until they run out by other mechanisms (in our
case, by the termination of the program after 20 time steps).

These experiments point to an important hypothesis, namely that
idle waves can survive for a long time on a non-noisy system,
but might be damped away by deliberate injection of noise, which is just
a collection of statistical, short-term delays. 
We will investigate this in Section~\ref{sec:Decay} below.

%%%%%%%%%%%%%%%%%%%%%%%%%%%%%%%%%%%%%%%%%%%%%%%%%%%%%%%%%%%%%%%%%%%%%%%%%%%

\subsection{Wave propagation speed}

\begin{figure}
		\centering
		\begin{subfigure}[t]{0.48\linewidth} %height
			\begin{tikzpicture}
			\put(-2.5,2.8) {\includegraphics[width=0.845\textwidth,height=0.098 \textheight]{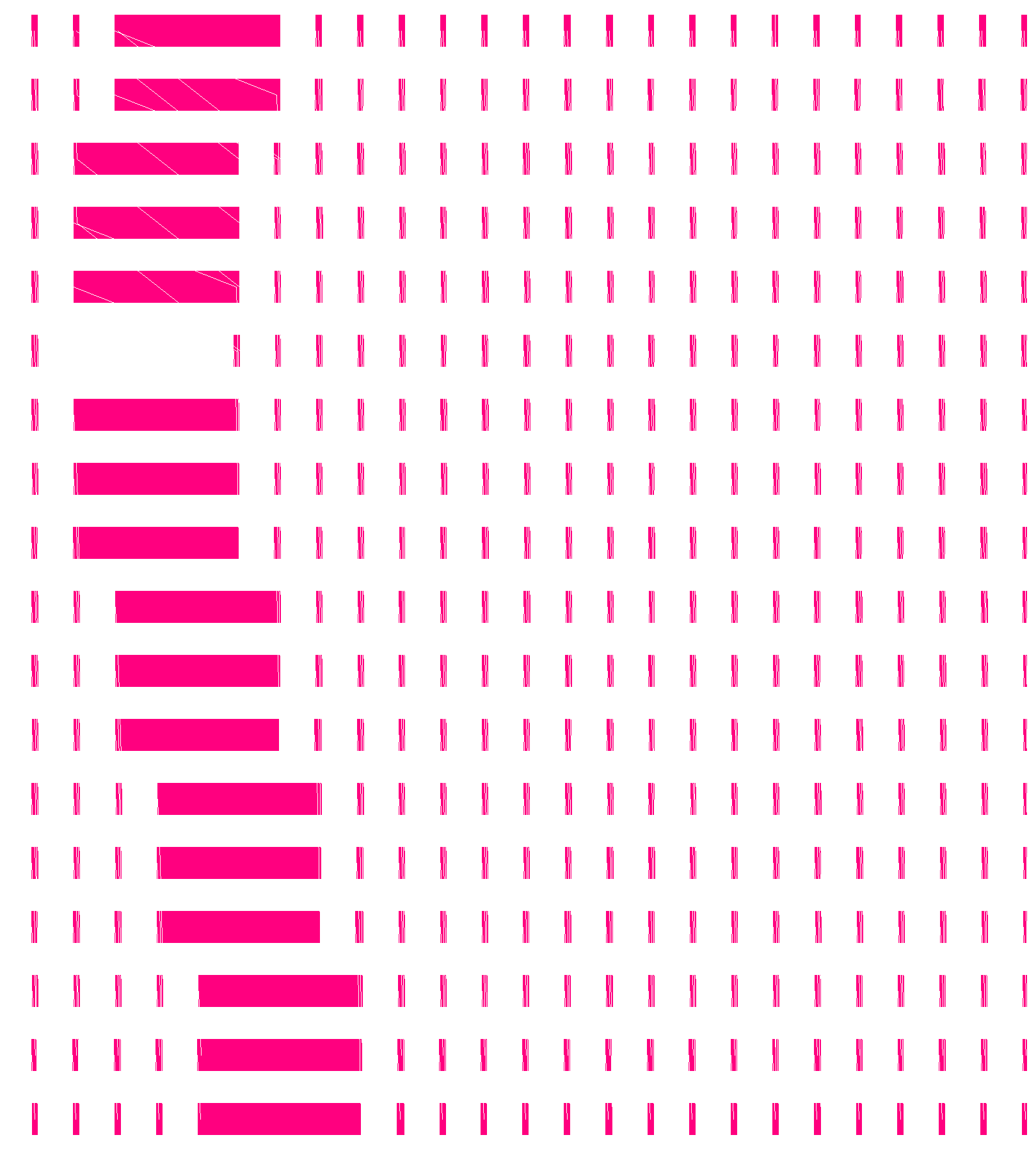}}
			\begin{axis}[
			width=1.188\textwidth,height=0.173 \textheight,
			ylabel = {Rank},
			y label style={at={(0.17,0.5)}},
			xlabel = {Time step},
			x label style={at={(0.5,0.08)}},
			x tick label style={font=\scriptsize},
			y tick label style={font=\scriptsize},
			xmin=1, xmax=24,
			ymin=0, ymax=19,
			xtick={1,2,3,4,5,6,7,8,9,10,11,12,13,14,15,16,17,18,19,20,21,22,23,24},
			ytick={1,2,3,4,5,6,7,8,9,10,11,12,13,14,15,16,17,18,19},
			xticklabels={,,2,,4,,,\textbf{4},,6,,8,,10,,12,,14,,16,,18,,20},
			yticklabels={17,,15,,13,,11,,9,,7,,5,,3,,1,},
			]
			\end{axis}
			\node [font=\small] at (1.5,-1.1){(a) Unidirectional}; 
			\node [font=\small] at (1.65,-1.5){open boundary rendezvous}; 
			\node [font=\small] at (0.45,1.71){\tikz \fill [blue] (1.57,0.1) rectangle (1,0.17);}; 
			\end{tikzpicture}
		\end{subfigure}
		\begin{subfigure}[t]{0.48\linewidth} %height
			\begin{tikzpicture}
			\put(-5.5,2.8) {\includegraphics[width=0.87\textwidth,height=0.098 \textheight]{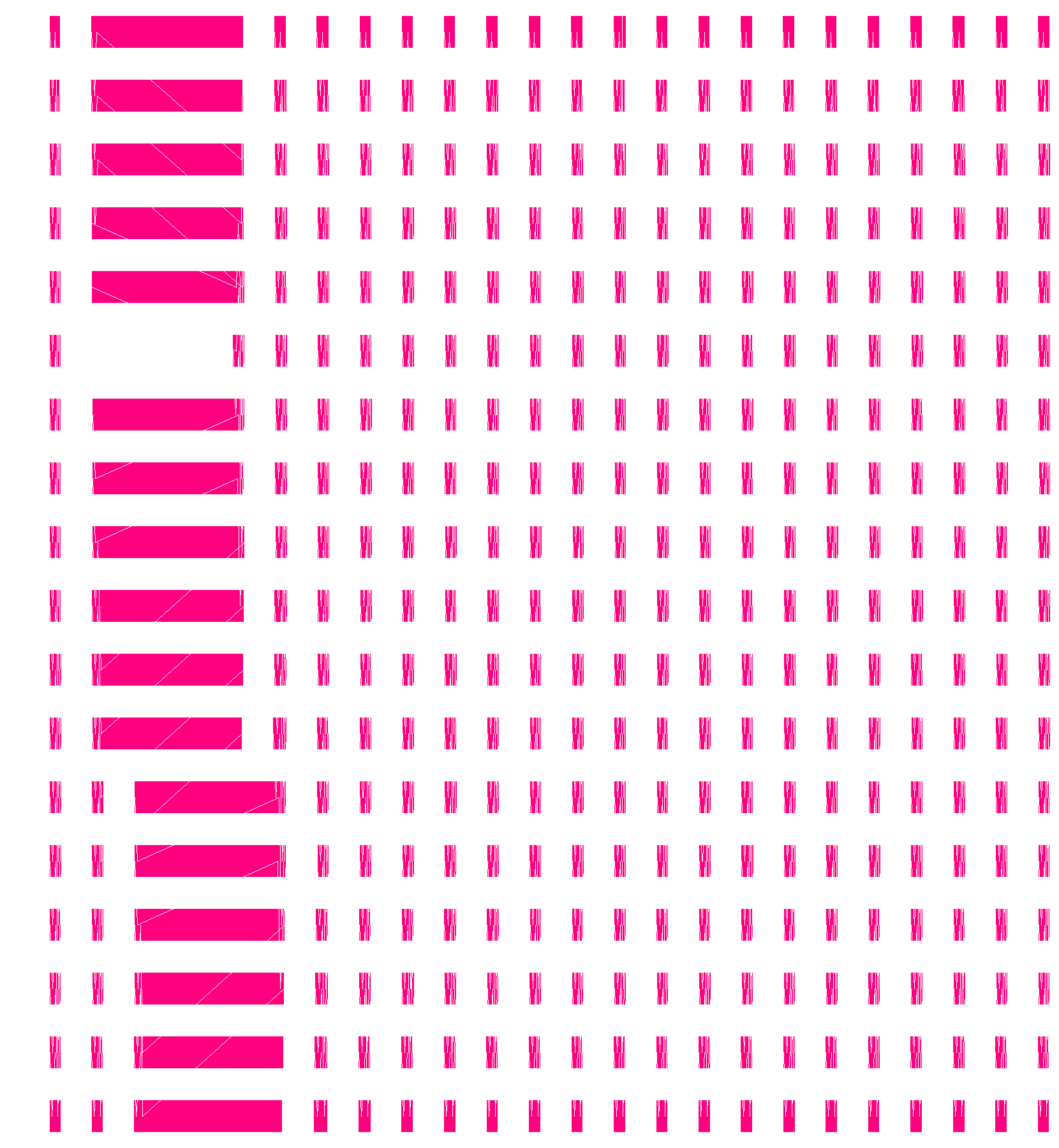}}
			\begin{axis}[
			width=1.188\textwidth,height=0.173 \textheight,
			ylabel = {Rank},
			y label style={at={(0.17,0.5)}},
			xlabel = {Time step},
			x label style={at={(0.5,0.08)}},
			x tick label style={font=\scriptsize},
			y tick label style={font=\scriptsize},
			xmin=1, xmax=24,
			ymin=0, ymax=19,
			xtick={1,2,3,4,5,6,7,8,9,10,11,12,13,14,15,16,17,18,19,20,21,22,23,24},
			ytick={1,2,3,4,5,6,7,8,9,10,11,12,13,14,15,16,17,18,19},
			xticklabels={,,2,,,\textbf{2},,4,,6,,8,,10,,12,,14,,16,,18,,20},
			yticklabels={17,,15,,13,,11,,9,,7,,5,,3,,1,},
			]
			\end{axis}
			\node [font=\small] at (1.5,-1.1){(b) Bidirectional}; 
			\node [font=\small] at (1.65,-1.5){open boundary rendezvous};
			\node [font=\small] at (0.4,1.71){\tikz \fill [blue] (1.535,0.1) rectangle (1,0.17);};  
			\end{tikzpicture}
		\end{subfigure}
	\caption{Timelines of delay propagation in a noise-free 
          environment with rendezvous protocol and next-to-next neighbor communication
          ($d=2$). Red color shows the propagating idle period (long bars) and communication delays
          (narrow bars). (a) Unidirectional communication, (b) bidirectional
          communication. }
	\label{fig:speed}
\end{figure}

Our experiments in Section~\ref{sec:Flavors} have shown that the speed of idle
wave propagation doubles with bidirectional rendezvous-mode communication
because the initial delay ``reaches out'' twice as far into neighboring
processes (Figures~\ref{fig:homo}(g), (h)). Basic analysis shows that,
on a noise-free system, this speed
depends in a simple manner on the execution period $T_\mathrm{exec}$, the
communication time $T_\mathrm{comm}$, the bidirectional rendezvous mode, and the
distance of neighbor communication $d$, which is the largest distance
to any communication partner of a process (up to now we have only
considered $d=1$):
\bq\label{eq:PropSpeed}
v_\mathrm{silent} = \frac{\sigma\cdot d}{T_\mathrm{exec}+T_\mathrm{comm}}~
\left[\frac{\mbox{ranks}}{\mbox{s}}\right]\cma
\eq
where
$$
\sigma = \left\{
\begin{array}{cl}
  2 & \mbox{~~if bidirectional rendezvous mode}\\ 
  1 & \mbox{~~if any other mode}
\end{array}
\right.\eos
$$ 
Note that it does not matter here what $T_\mathrm{comm}$ is composed of, be
it latency, overhead, transfer time, etc.. In fact, communication overhead and
execution time appear on an equal footing here. Figure~\ref{fig:speed} shows an
example with $d=2$ and rendezvous-mode unidirectional and bidirectional
communication, respectively: The presence of bidirectional communication
doubles the propagation speed. No such effect can be observed for
eager mode.

It turns out that even in a noisy system the propagation speed
along the ``forward,'' i.e., the leading slope of an idle wave
is hardly changed from $v_\mathrm{silent}$, while
the trailing slope is strongly influenced by it. The reason for this
is that system noise and past delays with all their accumulated
effects mainly interact with the trailing
edge of the idle wave. On any particular process, the delay
(i.e., the current manifestation of the idle wave) acts as a ``buffer''
and swallows much of the variation accumulated up to this
point. On the other hand, the idle wave can at most survive for one full
traversal of the process chain, so the interaction time of the leading edge
with noise is strictly limited.

The  effect of noise on the trailing edge of the wave
is investigated in the next section.

\section{Idle period decay}\label{sec:Decay}

\begin{figure}[tb]
	\begin{center}
		\begin{tikzpicture}
		\pgfplotstableread{figures/BackfrontDecay/meggie_E_MEDIAN_AVG_SD_MIN_MAX.txt}\meggiedata;		\pgfplotstableread{figures/BackfrontDecay/emmy_E_MEDIAN_AVG_SD_MIN_MAX.txt}\emmydata;
		\pgfplotstableread{figures/BackfrontDecay/Simulated_System.txt}\simulateddata;
		\begin{axis}[trim axis left, trim axis right, scale only axis,
		width = 0.80\columnwidth,
		height = 0.22\textheight,
		xlabel = {Mean delay, $E(\bar T^\mathrm{delay}_\mathrm{exec})$ [\%]},
		ylabel = {Average decay rate, $\bar\beta$ [\si{\micro \second \per rank}]},
		legend style = {nodes={inner sep=0.04em}, anchor=south, at={(0.27,0.75)}},
		legend columns = 1,
		]
		\addplot[only marks, mark=diamond*,  blue, error bars/.cd, y dir=both, y explicit,]
		table
		[
		x expr=\thisrow{E}, 
		y error minus expr=\thisrow{Median}-\thisrow{Min},
		y error plus expr=\thisrow{Max}-\thisrow{Median},
		]{\emmydata};
		\addplot[only marks, mark=square*,mark size =2 pt, Bittersweet, error bars/.cd, y dir=both, y explicit,]
		table
		[
		x expr=\thisrow{E}, 
		y error minus expr=\thisrow{Median}-\thisrow{Min},
		y error plus expr=\thisrow{Max}-\thisrow{Median},
		]{\meggiedata};
		\addplot+[only marks, mark=*, Fuchsia]
		table
		[
		x expr=\thisrow{E}, 
		y expr=\thisrow{DecayRate}
		]{\simulateddata};
       \legend{ InfiniBand system, Omni-Path system, Simulated system}
		\end{axis}
		\end{tikzpicture}
	\end{center}
	\caption{The average decay rate of an idle period in $\MuS$ per rank
          on the InfiniBand and Omni-Path clusters, and using the LogGOPSim
          simulator for comparison. Results show median,
          minimum and maximum values of statistics.}
	\label{fig:comparison}
\end{figure}
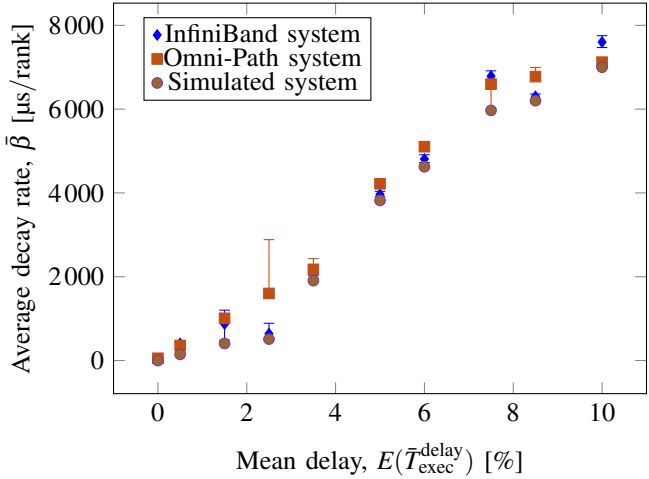

\subsection{Noise and decay rate}

It has been known for some time~\cite{markidis2015idle} that idle waves tend to
decay under the influence of system noise, but there was no quantitative
analysis so far. Here we analyze the average decay rate $\bar\beta$ of a single
idle wave while deliberately injecting fine-grained application noise
of an average length $\bar T^\mathrm{delay}_\mathrm{exec}$ into every execution
phase. This extra noise is exponentially distributed in order to mimic the
natural noise distribution on our systems (see Figure~\ref{fig:Systemnoise}).
It has the probability density function
\bq
f\left(\frac{T^\mathrm{delay}_\mathrm{exec}}{T_\mathrm{exec}};\lambda\right)=\lambda
\exp\left(-\lambda \frac{T^\mathrm{delay}_\mathrm{exec}}{T_\mathrm{exec}}\right)\eos
\eq
The parameter we use to characterize the noise is $E=\lambda^{-1}$
and quantifies the mean relative delay per execution period. 

Figure~\ref{fig:comparison} shows the measured decay rate with
statistics over 15 runs on our two cluster systems and, for
reference, a modified version of the LogGOPSim simulator (implementing
a simple Hockney model), versus the average delay ratio $E$.
There is
no qualitative difference among the three data sets, so the decay rate is
independent of the existing system noise. There is also a clear positive
correlation between the noise level and the decay rate, although more
measurements are required in order to be able to discern a definite functional
dependence.  Note that we chose our standard execution and communication
parameters as defined in Section~\ref{sec:Mechanics}. Unless
the idle wave is very narrow (incurring massive statistical variation),
the decay rate does not depend on the length of the injected delay.
For the above experiment we injected long delays of 90\,ms.

\subsection{Idle period elimination}

\begin{figure}[tb]
	\begin{subfigure}[t]{0.148\textwidth} %height
		\begin{tikzpicture}
		\put(0.5,5.8) {
		\includegraphics[width=0.915\textwidth,height=0.175 \textheight]{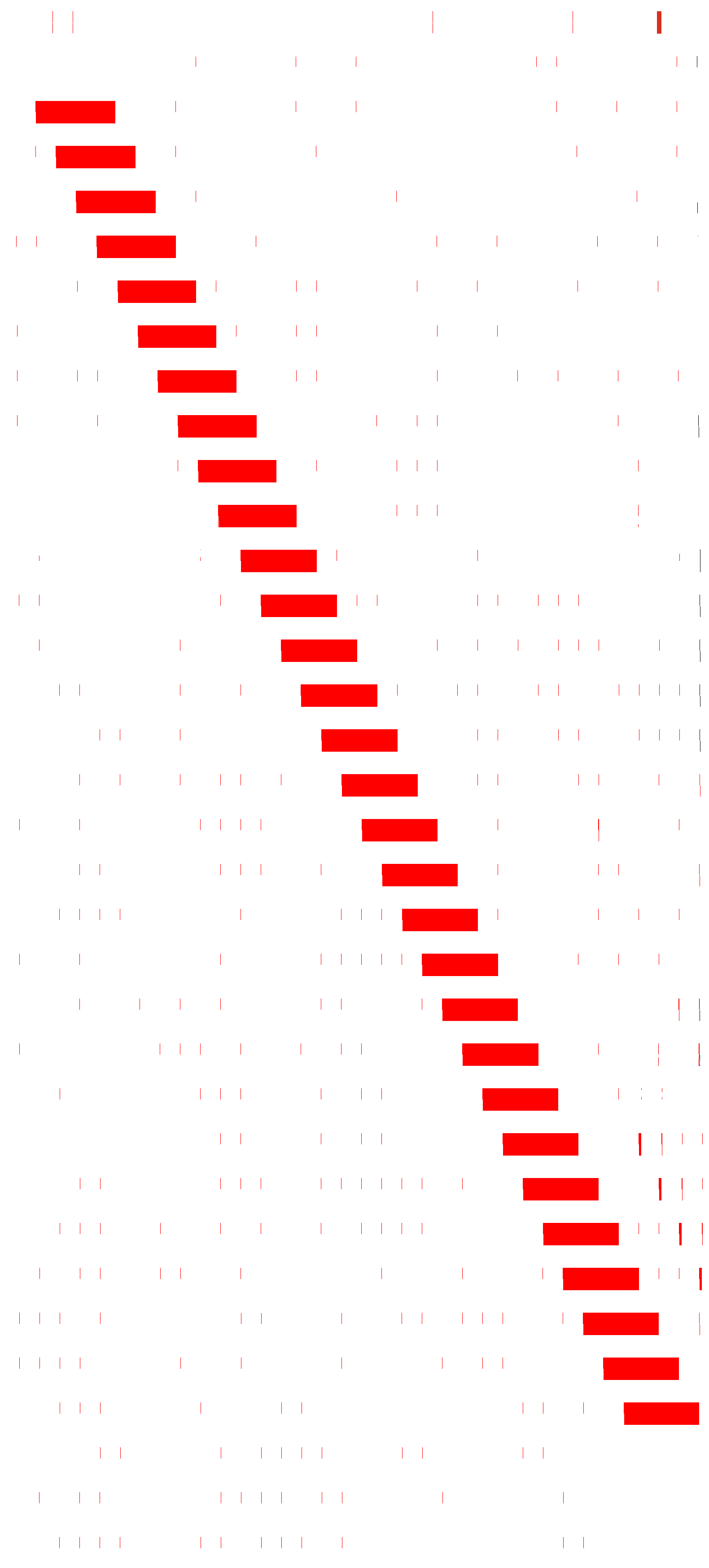}
	}
		\begin{axis}[
		width=1.5\textwidth,height=0.258\textheight,
		ylabel = {Rank},
		y label style={at={(0.17,0.5)}},
		xlabel = {Time step},
		x label style={at={(0.5,0.08)}},
		x tick label style={font=\scriptsize},
		y tick label style={font=\scriptsize}, 
		xmin=1, xmax=36,
ymin=0, ymax=38,
xtick={3,4,5,6,7,8,9,10,11,12,13,14,15,16,17,18,19,20,21,22,23,24,25,26,27,28,29,30,31,32},
		ytick={2,3,4,5,6,7,8,9,10,11,12,13,14,15,16,17,18,19,20,21,22,23,24,25,26,27,28,29,30,31,32,33,34,35,36}, 
		xticklabels={,,,,,,,,,,,,,,,,,,,,,,,,,,,,,30},
		yticklabels={\textbf{34},,,,,\textbf{29},,,,,\textbf{24},,,,,\textbf{19},,,,,\textbf{14},,,,,\textbf{9},,,,,\textbf{4},,,,},
		]
		\begin{scope}[on background layer]
		\fill[Apricot,opacity=0.2] ({rel axis cs:0.89,0}) rectangle ({rel axis cs:1,1});
		\fill[Apricot,opacity=0.5] ({rel axis cs:0.89,0.13}) rectangle ({rel axis cs:1,0.905});
		\end{scope}
		\end{axis}
		\node at (1,-1.1){(a) $E=0\%$}; 
		\node [font=\small] at (1.35,-1.5){$t_{total}=$\SI{51.1}{\milli \second}}; 
		\draw [semithick, dotted] (-0.6,3.68) -- (2.5,3.68); 
		\draw [semithick, dotted] (-0.6,2.97) -- (2.5,2.97); 
		\draw [semithick, dotted] (-0.6,2.255) -- (2.5,2.255); 
		\draw [semithick, dotted] (-0.6,1.545 ) -- (2.5,1.545 );  
		\draw [semithick, dotted] (-0.6,0.835 ) -- (2.5,0.835 ); 
	 \node [font=\small] at (0.275,4.15){\tikz \fill [blue] (0.1,0.1) rectangle (0.38,0.16);}; 
		\end{tikzpicture}
	\end{subfigure}
	\begin{subfigure}[t]{0.143\textwidth} %height
		\begin{tikzpicture}
		\put(2.5,5.8) {
	\includegraphics[width=0.884\textwidth,height=0.175 \textheight]{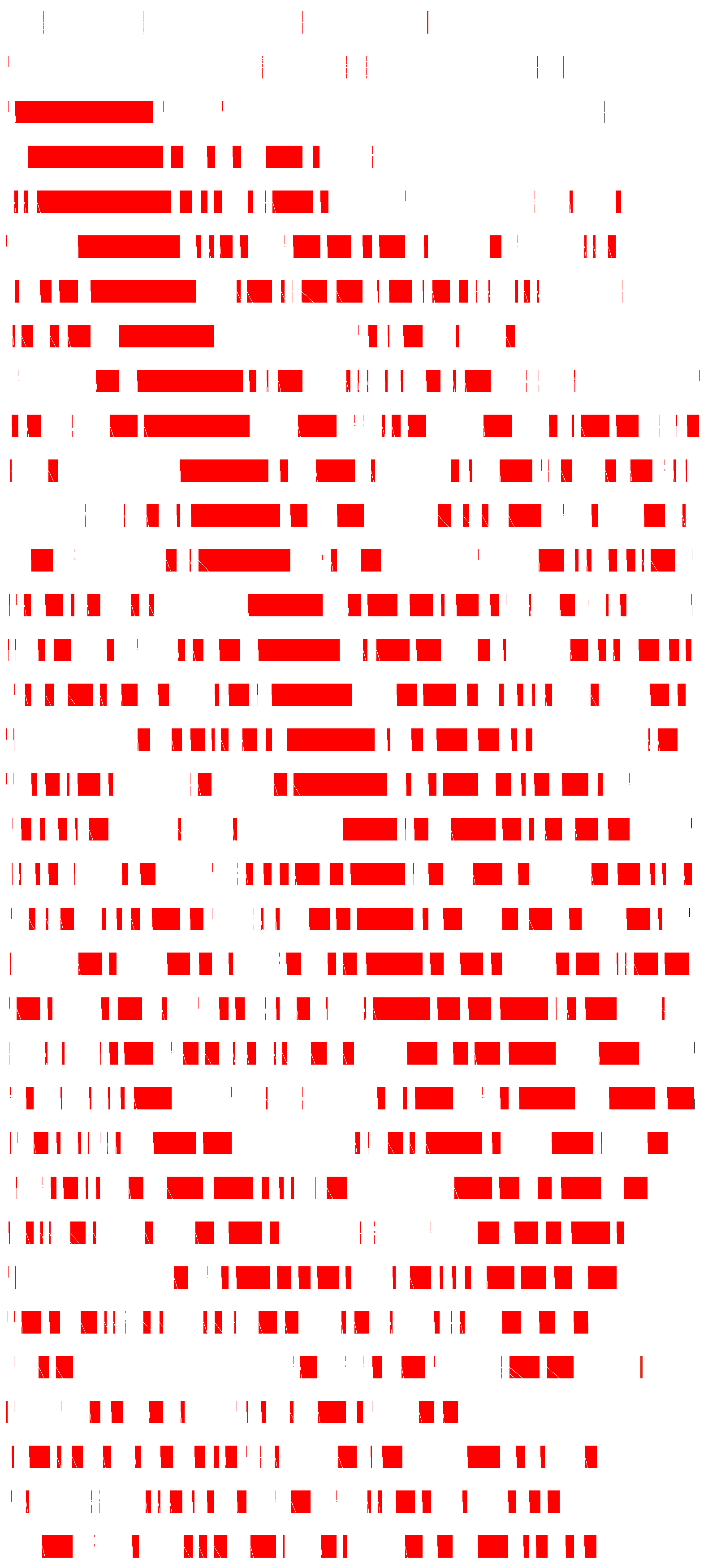}}
		\begin{axis}[
		width=1.55\textwidth,height=0.258\textheight,
			ylabel = {},
		y label style={at={(0.17,0.5)}},
		xlabel = {Time step},
		x label style={at={(0.5,0.08)}},
		x tick label style={font=\scriptsize},
		y tick label style={font=\scriptsize},
		xmin=1, xmax=36,
		ymin=0, ymax=38,
		xtick={3,4,5,6,7,8,9,10,11,12,13,14,15,16,17,18,19,20,21,22,23,24,25,26,27,28,29,30,31,32},
		ytick={2,3,4,5,6,7,8,9,10,11,12,13,14,15,16,17,18,19,20,21,22,23,24,25,26,27,28,29,30,31,32,33,34,35,36}, 
		xticklabels={,,,,,,,,,,,,,,,,,,,,,,,,,,,,,30},
		yticklabels={,,,,,,,,,,,,,,,,,,,,,,,,,,,,,,,,,,},
		]
		\begin{scope}[on background layer]
		\fill[Apricot,opacity=0.2] ({rel axis cs:0.89,0}) rectangle ({rel axis cs:1,1});
		\fill[Apricot,opacity=0.5] ({rel axis cs:0.89,0.26}) rectangle ({rel axis cs:0.965,0.72});
		\end{scope}
		\end{axis}
		\node at (1,-1.1){(b) $E=20\%$};  
		\node [font=\small] at (1.35,-1.5){$t_{total}=$\SI{82.7}{\milli \second}}; 
		\draw [semithick, dotted] (-0.3,3.68) -- (2.45,3.68); 
		\draw [semithick, dotted] (-0.3,2.97) -- (2.45,2.97); 
		\draw [semithick, dotted] (-0.3,2.255) -- (2.45,2.255); 
		\draw [semithick, dotted] (-0.3,1.545 ) -- (2.45,1.545 );  
		\draw [semithick, dotted] (-0.3,0.835 ) -- (2.45,0.835 );
        \node [font=\small] at (0.355,4.15){\tikz \fill [blue] (0.87,0.1) rectangle (0.4,0.16);}; 
		\end{tikzpicture}
	\end{subfigure}
	\begin{subfigure}[t]{0.143\textwidth} %height 
		\begin{tikzpicture}
		\put(2.5,5.8) {
	\includegraphics[width=0.871\textwidth,height=0.175 \textheight]{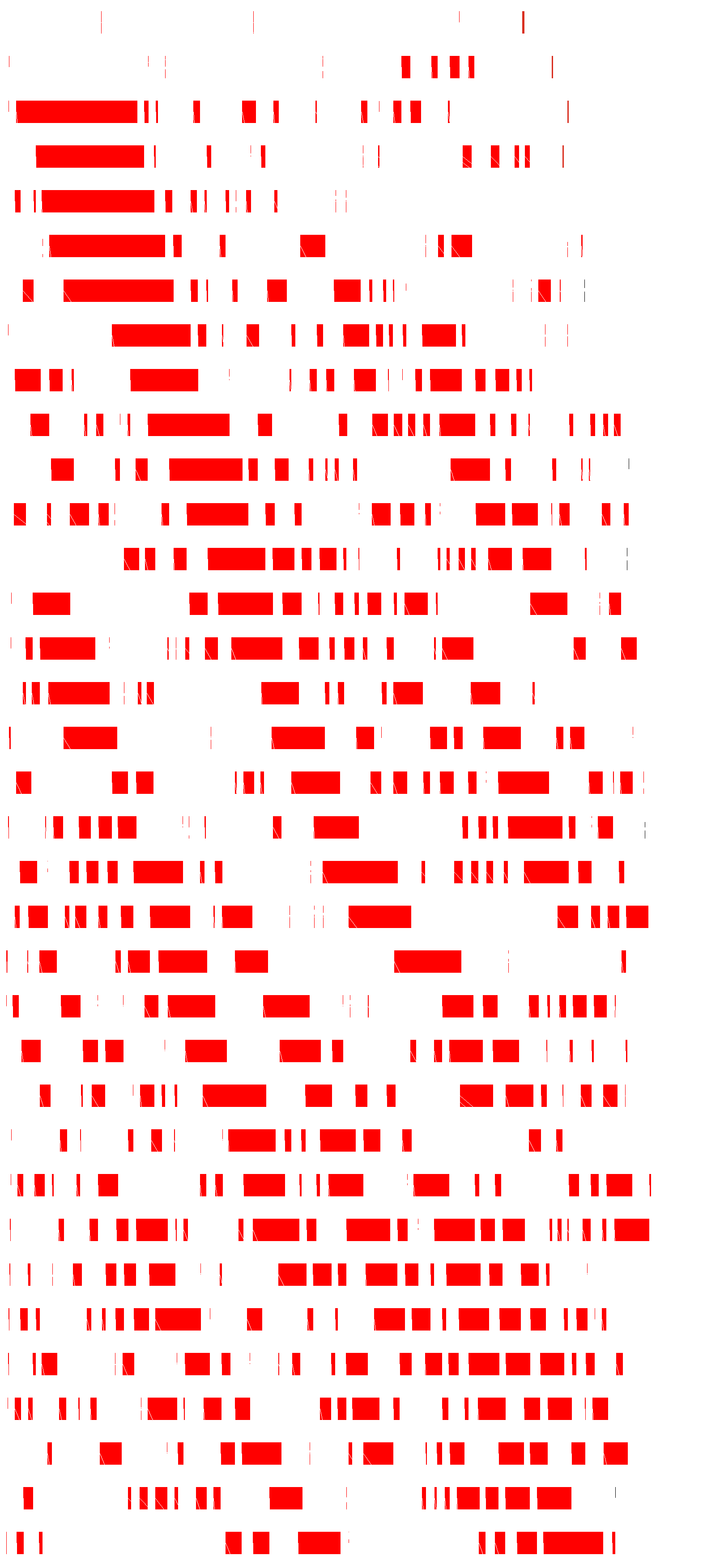}}
		\begin{axis}[
		width=1.55\textwidth,height=0.258\textheight,
			ylabel = {},
		y label style={at={(0.17,0.5)}},
		xlabel = {Time step},
		x label style={at={(0.5,0.08)}},
		x tick label style={font=\scriptsize},
		y tick label style={font=\scriptsize},
		xmin=1, xmax=36,
		ymin=0, ymax=38,
		xtick={3,4,5,6,7,8,9,10,11,12,13,14,15,16,17,18,19,20,21,22,23,24,25,26,27,28,29,30,31,32},
		ytick={2,3,4,5,6,7,8,9,10,11,12,13,14,15,16,17,18,19,20,21,22,23,24,25,26,27,28,29,30,31,32,33,34,35,36}, 
		xticklabels={,,,,,,,,,,,,,,,,,,,,,,,,,,,,,30},
		yticklabels={,,,,,,,,,,,,,,,,,,,,,,,,,,,,,,,,,,},
		]
		\begin{scope}[on background layer]
		\fill[Apricot,opacity=0.2] ({rel axis cs:0.89,0}) rectangle ({rel axis cs:1,1});
		\end{scope}
		\end{axis}
		\node at (1,-1.1){(c) $E=25\%$};
		\node [font=\small] at (1.35,-1.5){$t_{total}=$\SI{84.6}{\milli \second}}; 
		\draw [semithick, dotted] (-0.3,3.68) -- (2.5,3.68); 
		\draw [semithick, dotted] (-0.3,2.97) -- (2.5,2.97); 
		\draw [semithick, dotted] (-0.3,2.255) -- (2.5,2.255); 
		\draw [semithick, dotted] (-0.3,1.545 ) -- (2.5,1.545 ); 
		\draw [semithick, dotted] (-0.3,0.835 ) -- (2.5,0.835 ); 
		\node [font=\small] at (0.335,4.15){\tikz \fill [blue] (0.8,0.1) rectangle (0.4,0.16);}; 
		\end{tikzpicture}
	\end{subfigure}
	\caption{Damping of an idle wave by exponential
          noise of different average duration (zero, 20\%, and 25\%)
          on the InfiniBand cluster, running six processes per socket
          on six sockets (three nodes). An idle wave with a length of
          four execution periods (6\,ms) is injected at time step $1$ and rank $1$.
          Red color shows the sum of
          communication time and communication delays, and dotted lines
          denote socket boundaries.
        }
	\label{fig:DOM}
\end{figure}
Finally we investigate a core-bound parallel code under the influence
of an idle wave and variable noise. Figure~\ref{fig:DOM}(a) depicts the
noise-free situation (natural system noise is present but insignificant).
We show application time steps (up to 30) and, in addition, extra
wallclock time (orange bar) caused by the idle wave. 
After 30 time steps, the excess runtime is roughly equal to the injected delay
(6\,ms). With exponential noise injection at $E=20\%$ (Figure~\ref{fig:DOM}(b))
we can observe the strong decay of the idle wave but the execution time is only
marginally smaller. Of course, the overall runtime increases due to
the presence of noise. However, the processes causing the excess time are now
concentrated near the middle of the set. In Figure~\ref{fig:DOM}(c)
at $E=25\%$ we observe no excess runtime -- the idle period was damped
away by the noise.

\section{Related work}
\label{sec:Relatedwork}

Much interesting research has been conducted about the characterization
of system noise, its
impact on code performance, and how to mitigate
it~\cite{petrini2003case,jones2003improving,terry2004improving,gioiosa2004analysis,tsafrir2005system,beckman2006influence,ferreira2008characterizing,morari2011quantitative}.
However, little insight is available about how perturbations
of regular communication and communication structure travel through
and interact with a cluster system and the parallel applications
running on it.

The initial motivation for our work was provided by Markidis et
al.~\cite{markidis2015idle} and Peng et al.~\cite{peng2016idle}
who, based on results from the LogGOPSim
simulator~\cite{hoefler-loggopsim}, used Fourier analysis to learn
that isolated idle periods propagate among \ac{MPI} processes as
nondispersive linear waves. Their expression for propagation speed was
purely phenomenological, though, and missed the pivotal
ingredients of communication distance ($d$) and mode ($\sigma$),
which are part of our model (\ref{eq:PropSpeed}). This makes our
model a starting point for the investigation of collective
communication primitives. Their speculation that idle waves may be
described by a linear wave equation with damping cannot be
upheld, as our analysis of idle wave interaction shows.
Although they also observed the idle wave damping phenomenon, no quantitative
investigation of the connection between damping and noise was
provided.

\section{Conclusion and future work}
\label{sec:conclusions}

We have explored the phenomenology of idle waves in point-to-point
message-passing parallel programs with regular, bulk-synchronous
structure and core-bound performance characteristic communicating
via a non-blocking flat network infrastructure.  When an idle
wave, which typically initiated by a strong delay on one of the
processes, travels through the system, it does so with a certain speed
that depends on the range of point-to-point communication between
individual processes, the communication mode (eager vs.\ rendezvous)
and the direction of communication (one-directional
vs.\ bidirectional).  Idle waves interact by partial cancellation,
which indicates that a linear wave propagation model cannot be
applied.  This led to the hypothesis that fine-grained noise
may be capable of interacting with idle waves. Indeed we
have shown that in the presence of noise, the forward edge of the idle wave
is rather insensitive to the noise amplitude but its backward edge is
changed, leading to a damping effect. Running experiments with
exponentially distributed execution noise injections, we have observed
a clear positive correlation between the average noise amplitude
(relative to the undisturbed execution phase) and the decay rate of
the idle wave. We have further shown that the impact of idle waves
on the runtime of programs is limited on a noisy system, to the point
where the wave is completely absorbed by the noise.

We have only started to explore the huge parameter space of idle wave phenomena,
and our findings open possibilities for future research in many directions. Our
idle wave propagation model shows that the speed of idle waves depends on the
communication time per process, which can be different due to hierarchical
system structure (multicore CPUs, multiple sockets per node, network
topology). Hence, the propagation speed changes whenever a domain boundary is
crossed. This effect will be analyzed further. We will also look into code with
memory-bound characteristic (such as the motivating triad and LBM examples in
the introduction) because they bear a strong potential for desynchronization
and, thus, better utilization of the memory bandwidth and potential automatic
execution-communication overlap. In this context it will be useful to compare
pure MPI and hybrid MPI/OpenMP code since the latter tends to enforce frequent
thread synchronization, lessening the potential for inter-process skew.
Most of the examples in this paper were run in non-blocking communication
mode with a simple \verb.Isend./\verb.Irecv./\verb.Waitall. pattern.
We will explore how more advanced point-to-point and also
collective communication patterns influence the idle wave phenomenon.
Finally, our long-term goal is to establish a nonlinear continuum
model of message-passing programs that describes collective
phenomena like long-distance correlations and structure formation.

\ifblind
\else
\section*{Acknowledgments}
This work was supported by KONWIHR, the Bavarian Competence Network
for Scientific High Performance Computing in Bavaria, under project name ``OMI4papps.''
We are indebted to Thomas Zeiser and Michael Meier (RRZE) for
excellent technical support.
\fi

%\AtNextBibliography{\small}
\bibliographystyle{plain}
\bibliography{references}
%\printbibliography

\end{document}